\documentclass[letterpaper,floatfix,aps,prb,amsmath,amssymb,twocolumn,superscriptaddress,nopacs]{revtex4-2}

\usepackage{graphicx}
\usepackage{verbatim}
\usepackage{color}
\usepackage{array}
\usepackage[subnum]{cases}

\usepackage[pdfauthor={Fischer, Catelani},
            pdftitle={Nonequilibrium quasiparticle distribution in superconducting resonators: analytical approach}]{hyperref}

\hypersetup{pdfstartview={XYZ null null 1.05}}
\begin{document}

\title{Nonequilibrium quasiparticle distribution in superconducting resonators: \\ analytical approach}

\author{P. B. Fischer}

\affiliation{JARA Institute for Quantum Information (PGI-11), Forschungszentrum J\"ulich, 52425 J\"ulich, Germany}

\affiliation{JARA Institute for Quantum Information, RWTH Aachen University, 52056 Aachen, Germany}

\author{G. Catelani}

\affiliation{JARA Institute for Quantum Information (PGI-11), Forschungszentrum J\"ulich, 52425 J\"ulich, Germany}

\affiliation{Quantum Research Center, Technology Innovation Institute, Abu Dhabi 9639, UAE}

\begin{abstract}
In the superconducting state, the presence of a finite gap in the excitation spectrum implies that the number of excitations (quasiparticles) is exponentially small at temperatures well below the critical one. Conversely, minute perturbations can significantly impact both the distribution in energy and number of quasiparticles. Typically, the interaction with the electromagnetic environment is the main perturbation source driving quasiparticles out of thermal equilibrium, while a phonon bath is responsible for restoration of equilibrium. Here we derive approximate analytical solutions for the quasiparticle distribution function in superconducting resonators and explore the impact of nonequilibrium on two measurable quantities: the resonator's quality factor and its resonant frequency. Applying our results to experimental data, we conclude that while at intermediate temperatures there is clear evidence for the nonequilibrium effects due to heating of the quasiparticles by photons, the low-temperature measurements are not explained by this mechanism.
\end{abstract}

\date{\today}

\maketitle

\section{Introduction}
\label{sec:intro}

Nonequilibrium effects in superconductors have long attracted the interest of both experimentalists and theorists, starting with observations in the 1960s of enhancements in the critical current of weak links subjected to microwaves~\cite{Wyatt1966,Dayem1967}. Soon after, theoretical work~\cite{Eliashberg1970} predicted enhancements not only of the critical current, but also of the critical temperature $T_c$ and the gap $\Delta$ (see~\cite{Mooij1981} for an early review, and~\cite{Klapwijk2020} for a recent one). These effects are generally related to a redistribution in energy of quasiparticles: in a superconductor, the density of states is lower at higher energy, so a given number of quasiparticle excitations is less harmful to superconductivity if they are shifted to higher energy by the microwaves. The effects are more evident near $T_c$, where they were initially discovered, but more recently the focus has shifted to temperatures small compared to $T_c$, where nonequilibrium quasiparticles can be a resource or a complication. They are a resource in detectors such as kinetic inductance~\cite{Day2003} and nanowire single-photon detectors~\cite{Natarajan2012}, while they negatively affect qubits, electron pumps and turnstiles, and microrefrigerators~\cite{MQT2022}. 
A fundamental  question that we address here is how the quasiparticles redistribute in energy at low temperature $T\ll T_c$. The theoretical model to study this question can be written in the form of a kinetic equation for the quasiparticle distribution function in the presence of the microwave drive and accounting for the interaction with phonons, whose distribution is also determined by a kinetic equation~\cite{Chang.1977}. Solving for the distribution function in the presence of microwave is in general challenging, since the full model consists of coupled nonlinear integral equations. Assuming $T=0$, some basic properties of the solution were already considered in~\cite{Ivlev1973}; recently, a more detailed analysis of this case, where linearization is possible, has been given in~\cite{G.Catelani.2019}. The main qualitative result of that work is the identification of two regimes, cold and hot quasiparticles: cold quasiparticles have largely energy close to the gap, while hot ones are more broadly distributed over an energy $T_*$ which depends on the strength of the microwave drive; the latter determines the transition between the two regimes. The assumption $T=0$ means that the phonon distribution is fixed to be zero; to our knowledge, the full model allowing for nonequilibrium phonon has been studied only numerically~\cite{Goldie.2012}. Here we build on the results of Ref.~\cite{G.Catelani.2019} to arrive at an analytical description of the hot-quasiparticles case beyond the linearized regime. This description in turn enables us to derive explicit formulas for the internal quality factor of superconducting resonators that can be compared to experimental measurements. Our main finding is that the energy scale $T_*^3/\Delta^2$ separates two qualitatively different regimes: for phonon temperature above this scale, the quasiparticle density is close to its thermal value, the quality factor decreases exponentially with temperature and increases with drive strength; vice-versa, for temperature below this scale the density is much higher than the thermal value, the quality factor depends weakly on temperature and decreases with drive strength. For temperatures above $T_*^3/\Delta^2$ our results agree quantitatively with the measurements reported in Ref.~\cite{Visser.2014}.

Although the focus of this article is on the nonequilibrium quasiparticle distribution, our results are relevant to applications such as kinetic inductance detectors. For example, the existence of two different regimes depending on temperature and microwave drive strength could impact the way the detectors are characterized and their response calibrated~\cite{Gao.2008b}. This work can also enable further research on the impact of material parameters on the detector responsivity, see for instance Ref.~\cite{Valenti} and references there. More broadly, understanding the effects of phonons and photons on the quasiparticle distribution in resonators can provide a reference point in the study of other superconducting systems, such as nanobridge junctions~\cite{Levenson.2014} and granular superconductors~\cite{Gruenhaupt.2018}.

In the next section we briefly review  the kinetic equations for quasiparticles and phonons to establish our notation. In Sec.~\ref{sec: Shape of the quasiparticle distribution} we first extend the $T=0$ solution for the quasiparticle distribution to a wider energy range than that of~\cite{G.Catelani.2019}; then we consider finite phonon temperature as well as deviations of the phonon distribution from its equilibrium form. Going beyond the linearized model, in Sec.~\ref{sec: Quasiparticle Density} we consider the effect of the microwave photons on quasiparticle density and superconducting gap; the analytical results are validated by comparison with numerical solutions. Section~\ref{sec: Quality factors} presents the calculation of quality factor and resonant frequency, as well as comparison to experiments. We summarize our work in Sec.~\ref{sec:conclusions}.

\section{Kinetic equations}
\label{sec:Rate Equation}

The quasiparticle distribution function $f(E)$ in a superconductor obeys the kinetic equation
\begin{equation}\label{sec:Rate Equation eq: Rate Equation}
\frac{df(E)}{dt}=St^{Phon}\{f,n\}+St^{Phot}\{f,\bar{n}\}
\end{equation} 
with the two collision integrals $St^{Phon}\{f,n\}$ and $St^{Phot}\{f,\bar{n}\}$ accounting for the interaction of quasiparticles with phonons and photons, respectively. In our notation, $n(\omega)$ represents the distribution function of phonons and $\bar{n}$ the (average) number of photons. The collision integrals can be derived using non-equilibrium Green's functions~\cite{G.MEliashberg.1972} or Fermi's golden rule~\cite{Chang.1978}, and can be generally split into terms that conserve or change the number of quasiparticles. Note that we assume the system to be homogeneous, so the distribution functions are independent of position. The kinetic equation is complemented by the self-consistent equation for the superconducting gap $\Delta$
\begin{equation}\label{sec:Rate Equation eq: Gap Equation}
\ln \left( \frac{\Delta_0}{\Delta}
\right)=\int\limits^{\infty}_{\Delta} dE \, \rho(E) \frac{2f(E)}{E} 
\, ,
\end{equation}
where $\Delta_0$ is the zero-temperature gap (that is, in the absence of quasiparticles) and
\begin{equation}\label{sec:Rate Equation eq:  DOS}
\rho(E)=\frac{E}{\sqrt{E^2-\Delta^2}}
\end{equation}
is the normalized density of states.
 
\subsection{Interaction with phonons}

For the phonon collision integral, in the term conserving the quasiparticle number we distinguish spontaneous emission of phonons from stimulated emission and absorption, while the number non-conserving terms account for recombination of quasiparticles into Cooper pairs and pair-breaking events:
\begin{equation}\label{sec:Rate Equation eq: Phonon integral}
\begin{split}
St^{Phon}\{f,n\}=&St^{Phon}_{sp}\{f\}+St^{Phon}_{st}\{f,n\}\\
+&St^{Phon}_r\{f,n\}+St^{Phon}_{PB}\{f,n\}.
\end{split}
\end{equation}
The first term on the right hand side describes spontaneous emission (we use units with $k_B=\hbar=1$):
\begin{equation}
\begin{split}\label{sec:Rate Equation eq: Gamma^Phon_sp}
&St^{Phon}_{sp}\{f\} =\frac{1}{\tau_0 T_c^3} \\
&\times\bigg\{\int\limits_{0}^{\infty} \!d\omega\,\omega^2  U^- (E,E+\omega) f(E+\omega)\left[1-f(E)\right]\\
&-\int\limits_{0}^{E-\Delta} \!\!d\omega\,\omega^2 U^- (E,E-\omega)f(E) \left[1-f(E-\omega)\right]\bigg\}
\end{split}
\end{equation}
where $T_c \simeq \Delta_0/1.764$ is the critical temperature, and the functions
\begin{equation}\label{eq:Udef}
U^{\pm}(E,E')=\rho(E')K^{\pm }(E,E')    
\end{equation}
are given by a product between the density of states, Eq.~(\ref{sec:Rate Equation eq:  DOS}),
and the BCS coherence factors~\cite{Bardeen.1957} 
\begin{equation}\label{sec:Rate Equation eq: Coherence Factors}
K^{\pm}(E,E')=1\pm \frac{\Delta^2}{EE'}\, .
\end{equation}
Finally, the factor $\omega^2/T_c^3\tau_0$ accounts for the strength of the electron-phonon interaction using the Debye model and in a low-frequency approximation appropriate for weakly-coupled superconductors~\cite{Kaplan.1976}.

With the notation introduced above, the contribution of stimulated emission and absorption is
\begin{equation}\label{sec:Rate Equation eq: Stimulated Phonon integral}
\begin{split}
&St^{Phon}_{st}\{f,n\}=\frac{1}{\tau_0 T_c^3} \int\limits_{0}^{\infty}\!d\omega\,\omega^2  U^- (E,E+\omega) \\
&\times\left\{f(E+\omega)\left[1-f(E)\right]-f(E)\left[1-f(E+\omega)\right] \right\} n(\omega)\\
&+\frac{1}{\tau_0 T_c^3} \int\limits_{0}^{E-\Delta} \!d\omega\,\omega^2 U^- (E,E-\omega) \\
&\times\left\{f(E-\omega)\left[1-f(E)\right]-f(E) \left[1-f(E-\omega)\right]\right\} n(\omega).
\end{split}
\end{equation}
Quasiparticle number conservation follows from the identities
\begin{equation}
    \int\limits_{\Delta}^{\infty} dE \, \rho(E) St_{sp}^{Phon} =\int\limits_{\Delta}^{\infty} dE \, \rho(E) St_{st}^{Phon} = 0  
\end{equation}
The recombination and pair-breaking terms are, respectively,
\begin{align}
St^{Phon}_r\{f,n\}=&-\frac{1}{\tau_0  T_c^3} \int\limits_{E+\Delta}^{\infty}\!d\omega\, \omega^2 U^+ (E,\omega-E) \nonumber \\ 
&\times f(E)f(\omega-E)\left[1+n(\omega)\right]
\label{sec:Rate Equation eq: Gamma^Phon_re}
\end{align}
and
\begin{align}
St^{Phon}_{PB}\{f,n\}=&\frac{1}{\tau_0  T_c^3} \int\limits_{E+\Delta}^{\infty} \!d\omega\, \omega^2 U^+ (E,\omega-E) \nonumber \\
&\times [1-f(E)]\left[1-f(\omega-E)\right]n(\omega).
\label{sec:Rate Equation eq: Gamma^Phon_PB}
\end{align}

\begin{figure*}[!bt]
    \centering
    \includegraphics[width=0.95\linewidth]{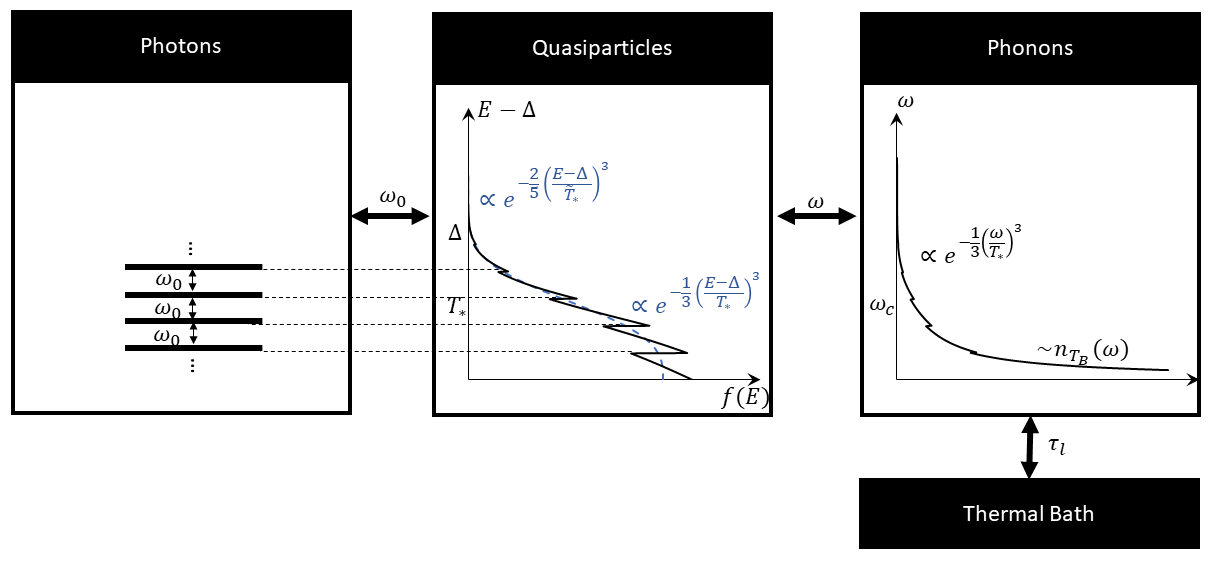}
    \caption{Schematic representation of the interaction of quasiparticles with photons and phonons. Absorption and emission of photons of frequency $\omega_0$ drives the quasiparticle distribution out of equilibrium. The divergence in the BCS density of states leads to a quasiparticle distribution with peaks at multiples of the photon energy above the gap; the peaks' amplitudes follow a slowly varying (on the scale $\omega_0$) envelope. The quasiparticles also exchange energy $\omega$ with the phonons in the superconductor;
    the phonons are in contact with a thermal bath, so their distribution can relax towards equilibrium over time $\tau_l$.} 
    \label{fig:System}
\end{figure*}

Since the processes described by $St^{Phon}$ involve the emission or absorption of a phonon, in a general non-equilibrium situation one must also consider the kinetic equation for the phonon distribution function~\cite{Chang.1977,Chang.1978},
\begin{align}
& \frac{d n(\omega)}{dt}=\frac{2 }{\pi \Delta_0 \tau^{PB}_0}\int\limits_{\Delta}^{\infty}\!dE\, \rho(E) U^-(E,E+\omega)\big\{ f(E+\omega) \nonumber\\
&\times\left[1-f(E)\right]\left[1+n(\omega)\right]-f(E)\left[1-f(E+\omega)\right] n(\omega) \big\} \nonumber\\
&+\frac{1 }{\pi \Delta_0 \tau^{PB}_0}\int\limits_{\Delta}^{\omega-\Delta}\!dE\, \rho(E) U^+(E,\omega-E)\big\{ f(\omega-E) \nonumber\\
&\times f(E)\left[1+n(\omega)\right]-\left[1- f(\omega-E)\right]\left[1- f(E)\right]n(\omega)\big\}\nonumber\\
&-\frac{1}{\tau_l}\left[n(\omega)-n_T(\omega,T_{B})\right] \,. 
\label{sec:Rate Equation eq: Phonon Rate Equation}
\end{align}
The first integral on the right-hand side is the counterpart to the quasiparticle-conserving collision integrals $St^{Phon}_{sp}$ and $St^{Phon}_{st}$ of Eqs.~(\ref{sec:Rate Equation eq: Gamma^Phon_sp}) and (\ref{sec:Rate Equation eq: Stimulated Phonon integral}), while the second integral to the recombination and pair-breaking terms of Eqs.~(\ref{sec:Rate Equation eq: Gamma^Phon_re}) and (\ref{sec:Rate Equation eq: Gamma^Phon_PB}), the factor of two in front of the first integral accounting for spin.
For frequencies $\omega<2\Delta$ the second integral has to be replaced by zero. Within a phenomenological relaxation time approach, the last term in Eq.~\eqref{sec:Rate Equation eq: Phonon Rate Equation} takes into account phonon exchange with a thermal equilibrium bath of temperature $T_B$, $n_T(\omega,T_B)= (e^{\omega/T_B}-1)^{-1}$. 
Note that 
the lifetime of a phonon of energy $2\Delta_0$ against pair breaking at zero temperature $\tau^{PB}_0$ and the characteristic time $\tau_0$ are related~\cite{Goldie.2012,Zehnder.1995},
\begin{equation}\label{sec:Rate Equation eq: Relation Lifetimes}
\frac{\tau_0}{\tau^{PB}_0}=\frac{2 \Delta_0 \pi \rho_F \omega_D^3}{9 N_\mathrm{ion} T_c^3}\, ,
\end{equation}
where $\rho_F$ is the single-spin electronic density of states at the Fermi energy, $\omega_D$ is the Debye frequency, and $N_\mathrm{ion}$ the ionic volume density. For Al, using the parameters reported in Ref.~\cite{Kaplan.1976} this gives $\tau_0/\tau^{PB}_0\simeq 1.7 \times 10^3$, but this ratio is  smaller for other materials considered there (for instance, it is about 36 for Nb).

\subsection{Interaction with photons}

For the photon collision integral, we consider a single mode of frequency $\omega_0<2\Delta$, so that no photon-mediated recombination or pair breaking can take place. 
Then the collision integral resembles the number-conserving contribution to the phonon collision integral: 
\begin{align}
& St^{Phot}\{f,\bar{n}\}=c_{Phot}^{QP} U^+(E,E+\omega_{0})\Big\{f(E+\omega_0)  \nonumber\\
&\times \left[1-f(E)\right] 
(\bar{n}+1)-f(E)\left[1-f(E+\omega_0)\right]\bar{n})\Big\} \nonumber\\
& +c_{Phot}^{QP} U^+(E,E-\omega_{0})\Big\{ f(E-\omega_0) \left[1-f(E)\right]\bar{n} \nonumber\\
& -f(E)\left[1-f(E-\omega_0)\right] (\bar{n}+1)
\Big\} 
\label{sec:Rate Equation eq:  Photon integral}
\end{align}
and is itself number-conserving, $\int_\Delta dE\, \rho(E) St^{Phot} = 0$.
The term in the second curly brackets is set to zero for $E-\Delta<\omega_0$. Here the average photon number $\bar{n}$ is treated as a known, independent quantity. More generally, it can be affected by the properties of the resonator, and in Sec.~\ref{sec: Quality factors} we will calculate $\bar{n}$ as function of the read-out power for a half-wavelength resonator coupled to a transmission line.
Following Ref.~\cite{G.Catelani.2019}, we define an effective temperature $T_0$ via $e^{\omega_{0}/T_0}\equiv (\bar{n} + 1)/\bar{n}$.
However, instead of approximating the density of states and coherence factors by their form near the gap, as done in Ref.~\cite{G.Catelani.2019}, we keep their full form, and in contrast to Refs.~\cite{Chang.1977,Chang.1978} we include spontaneous photon emission.
The coupling constant in resonators can be estimated as~\footnote{To check this expression, we note that the power $P_\mathrm{abs}$ absorbed by the quasiparticles can be calculated by multiplying $St^{Phot}$, Eq.~\eqref{sec:Rate Equation eq:  Photon integral}, by $4\rho_F V E \rho(E)$ and integrating over $E$; for $\bar{n}\gg1$, comparing the result to Eq.~\eqref{sec:Quality factors eq: sigma_1} for $\sigma_1$, we find $P_\mathrm{abs}=2\bar{n}\omega_0^2 c^{QP}_{Phot}\sigma_1/\sigma_N\delta$. The inverse quality factor is by definition $1/Q=P_\mathrm{abs}/\omega_0 \mathcal{E}$ with $\mathcal{E}=\bar{n}\omega_0$ the energy stored in the resonator [cf. Eq.~\eqref{sec: Quality factors eq: nbar}]. Using that in the normal state $\sigma_1=\sigma_N$, we arrive at Eq.~\eqref{Resonator_Q}.}
\begin{equation}\label{Resonator_Q}
c_{Phot}^{QP}=\frac{\delta}{2 Q'}
\end{equation}
with $Q'$ the quality factor of the resonator if the material resistivity were as in the normal state and $\delta\approx 1/V\rho_F$ the mean level spacing, where $V$ is the volume occupied by the quasiparticles. A schematic representation of the system under consideration is presented in Fig.~\ref{fig:System}.

\subsection{Numerical approach}
\label{sec:Rate Equation eq: Numerical approach}

The equations \eqref{sec:Rate Equation eq: Rate Equation} to  \eqref{sec:Rate Equation eq: Phonon Rate Equation} constitute a  system of coupled non-linear integral equations whose solution, even in the steady state, is clearly non-trivial. Here we describe briefly how we solve the system numerically.
To discretize the system 
in the steady state, we divide the energy axis in intervals  $\Xi_i=[\psi_i,\psi_{i+1}]$ with with $\psi_i=ih$ and average over each such intervals, so that the steady-state condition for Eq.~\eqref{sec:Rate Equation eq: Rate Equation} becomes
\begin{equation}
    \frac{d}{dt}\int\limits_{\Delta+\psi_i}^{\Delta+\psi_{i+1}} dE f(E) =0
\end{equation}
We choose the discretization such that the gap is an integer multiple of $h$, and additionally round the photon energy to the nearest integer multiple of $h$. Furthermore, for $E_i\in \Xi_i+\Delta$ and $\omega_j\in \Xi_j$ we approximate $f(E_i)=f(\Delta+\psi_i)$, $n(\omega_j)=n(\psi_j)$, $K^\pm(E_i,E_i\pm \omega_j)=K^\pm(\Delta+\psi_i,\Delta+\psi_{i\pm j})$ to convert Eq.~\eqref{sec:Rate Equation eq: Rate Equation} to a system of ordinary equations. This procedure is equivalent to replacing the quasiparticle density of states by the density of states averaged over each interval $\Xi_i$ and replacing the other quantities by their value on a grid as described above. A similar discretizetion procedure has been used in Ref.~\cite{Kozorezov.2004}; the major differences to our work are that we discretize phonon and quasiparticle kinetic equations separately and neglect the variation of the coherence factors $K^{\pm}$ and the $\omega^2$ factor over each interval, enabling us to calculate the weights analytically~\footnote{In contrast to our procedure, in which the BCS density of states Eq.~\eqref{sec:Rate Equation eq:  DOS} is kept unaltered, some authors introduce a broadened density of states and a cutoff using a Heaviside function \cite{Goldie.2012,Guruswamy.2018}. This approach can be problematic; for example, to our understanding the discretized version of the photon integral used in Ref.~\cite{Guruswamy.2018} violates the conservation of the number of quasiparticles. With our approach, introducing a broadening and a cutoff is not necessary, and terms that conserve the number of quasiparticles in Eq.~\eqref{sec:Rate Equation eq: Rate Equation} also conserve the number of quasiparticles in the discretized version.}. 
We use the same approach of replacing the quasiparticle density of states by the averaged one, and discretizing the other quantities as described above, to obtain the discretized version of Eq.~\eqref{sec:Rate Equation eq: Phonon Rate Equation}.

To arrive at the numerical solution of the full system, we proceed via intermediate steps that are similar to those we will employ to find approximate analytical results, see Secs.~\ref{sec: Shape of the quasiparticle distribution} and \ref{sec: Quasiparticle Density}. To begin with, we take the phonon distribution to be the thermal equilibrium one, and approximately keeping only terms linear in $f \ll 1$, Eq.~\eqref{sec:Rate Equation eq: Rate Equation} reduces to a matrix equation
\begin{equation}\label{sec:Rate Equation eq: matrix equation}
    \mathbf{M}f^0=0 \, ,
\end{equation}
where superscript 0 is used to denote the initial solution (that is, step zero) for the quasiparticle distribution function.
This equation can be solved by diagonalizing $\mathbf{M}$
but it does not determine the distribution's normalization. The latter is fixed by the non-linear terms, and its approximate value can be found using the discretized version of the approach described in Sec.~\ref{sec: Quasiparticle Density}.

After obtaining the properly normalized $f^0$, non-linear terms in Eq.~\eqref{sec:Rate Equation eq: Rate Equation} as well as Eq.~\eqref{sec:Rate Equation eq: Phonon Rate Equation} can be taken into account using Newton's algorithm. To solve the two equations simultaneously, we follow Ref.~\cite{Goldie.2012} by forming the state vector $x=(f,n)^T$ and writing the system of equations determining the state vector in the steady state in the form $g(x)=0$. Successive approximations $x^i$ to the state vector are calculated using the recursive relation
\begin{equation}\label{sec:Rate Equation eq: Newton Algorithm}
    x^{i+1}=x^i-J^{-1}(x^{i}) g(x^{i})
\end{equation}
with $J^{-1}$ the inverse of the Jacobian matrix of $g$.  The normalized $f^0$ is used as the initial guess for the quasiparticle distribution, while the initial guess for the phonon distribution can be obtained by inserting $f^0$ in Eq.~\eqref{sec:Rate Equation eq: Phonon Rate Equation}, setting the temporal derivative to zero, and solving that linear equation for $n$. Calculations of the initial phonon distribution and of the Jacobian matrix can be carried out analytically, thus avoiding long computational times and rounding errors.

The discretized system of equation is not differentiable with respect to the gap $\Delta$, so we do not include the self-consistency condition Eq.~\eqref{sec:Rate Equation eq: Gap Equation} in Newton's algorithm. To find deviations of the gap from its equilibrium value, we can solve Eq.~\eqref{sec:Rate Equation eq: Newton Algorithm} for a fixed gap. This gives the distribution $f$ as function of the gap, and we can use the result in a bisection algorithm applied to Eq.~\eqref{sec:Rate Equation eq: Gap Equation} to calculate the non-equilibrium gap. While this method gives a fully self-consistently calculated gap, it is accurate only up to variations of the gap of order $h$, as both the gap and the photon energy have to be rounded to integer multiple of $h$ in the calculation of the distribution function. For the small deviation $\delta= \Delta_0 - \Delta$ of the gap from its zero-temperature value encountered in this paper, we instead proceed as follows: the left-hand side of Eq.~\eqref{sec:Rate Equation eq: Gap Equation} is approximately $\delta/\Delta_0$ at leading order; therefore, in an iterative approach to solving the equation, we calculate
$f$ with the gap fixed at its thermal equilibrium value, thus neglecting the deviation of the gap from equilibrium in the right-hand side.
This gives an expression for $\delta$ in terms of an integral which depends only on the quasiparticle distribution obtained using the thermal equilibrium gap and that can be evaluated numerically. The numerical solution gives the quasiparticle distribution at points on an energy grid of spacing $h$; since we want to use a finer grid to numerically evaluate the integral, between these points we linearly interpolate the  distribution function (the interpolation can be used because the distribution function is smooth between two nearby peaks at multiples of $\omega_0$ above the gap).

\subsection{Relevant timescales}
\label{sec:Lifetimes}

While a numerical solution to the system of coupled integral equations, Eqs.~\eqref{sec:Rate Equation eq: Rate Equation} and \eqref{sec:Rate Equation eq: Phonon Rate Equation}, can be found as just described, an exact analytical solution is likely impossible except in particular cases such as thermal equilibrium; it is therefore instructive to discuss the different timescales governing the dynamics of the quasiparticles and phonons and under which conditions approximate analytical solutions might be found. 
Throughout this work we assume the quasiparticles to be non-degenerate, $f(E) \ll 1$, so we always neglect Pauli Blocking factors by replacing $[1-f(E)] \to 1$.

There are several characteristic times that can be read off from the kinetic equations, as detailed in Appendix~\ref{appendix:Lifetimes}. These lifetimes in general depend on the energy $E$ of the quasiparticle or $\omega$ of the phonon; we begin by discussing the phonon lifetimes. For phonons with energy above the pair-breaking threshold, $\omega > 2\Delta$, their pair-breaking lifetime $\tau^{Phon}_{PB}(\omega)$ depends weakly on energy near the threshold, so we take approximately $\tau^{Phon}_{PB}(\omega) \approx \tau_0^{PB}$; 
in Al, $\tau_0^{PB}\simeq 240\,$ps \cite{Kaplan.1976}. The lifetime due to the phonon being absorbed by a quasiparticle is in general much longer than this, $\tau^{Phon}_{abs}(\omega) \gg \tau_0^{PB}$, due to the assumed non-degeneracy (and hence low density) of the quasiparticles. In contrast, in aluminum films the thermalization time $\tau_l$ [last term in Eq.~\eqref{sec:Rate Equation eq: Phonon Rate Equation}] is comparable to the pair-breaking time, $\tau_l \sim \tau_0^{PB}$; for films of thickness $100\,$nm on a sapphire substrate, the thermalization time is estimated to be $\tau_l\simeq 500\,$ps~\cite{Kaplan.1979,Eisenmenger.1976,Chang.1978}.
Since thermalization and pair-breaking times are comparable, we expect them to have both an impact on the above-threshold phonon distribution and hence on the quasiparticle number. Moreover, the absorption time being long means that the phonon distribution can significantly deviate from the equilibrium one; however, as we discuss below, generally this does not impact the shape of the quasiparticle distribution. We note that these consideration can be material-specific; for instance, for a 100~nm thick Nb film on sapphire we estimate $\tau_{0}^{PB}\simeq 4\,$ps and $\tau_l\simeq 1.5\,$ns~\cite{Kaplan.1976,Kaplan.1979}, so deviations from equilibrium can become more significant compared to aluminum.

Turning to the quasiparticle lifetimes, we can identify two scattering times, $\tau^{qp}_{s,t}(E)$ and $\tau^{qp}_{s,n}(E)$, involving photons and phonons respectively and accounting for all possible number-conserving processes (absorption as well as spontaneous and stimulated emission), and the recombination lifetime $\tau^{qp}_r(E)$, which is inversely proportional to the quasiparticle density. In this work we focus on quasiparticles of energies up to a few times the superconducting gap, as the energy dependence of the phonon scattering time implies that quasiparticles of higher energies typically relax very quickly towards the gap and therefore do not contribute directly to processes like photon absorption. When the recombination lifetime is longer than the scattering ones, $\tau^{qp}_r \gg \tau^{qp}_{s,n},\, \tau^{qp}_{s,t}$, as is the case for low quasiparticle densities and sufficiently high photon number, the shape of the quasiparticle distribution function is determined by the number-conserving processes, while generation and recombination affect its normalization (that is, the overall quasiparticle density). Moreover, for strong deviations from equilibrium to be possible, the photon scattering time should be the shortest time scale, $\tau^{qp}_{s,t} \lesssim \tau^{qp}_{s,n}$, a condition that can be met if the number of photons $\bar{n}$ is sufficiently large and/or the phonon bath temperature $T_B$ sufficiently low; in particular, the phonon scattering time being longer than the photon one means that deviations of the phonon distribution from equilibrium have a small effect on the shape of the quasiparticle distribution (see also Sec.~\ref{sec:Shape of the quasiparticle distribution Phonon trapping}). 
Therefore in the next section we will study the shape of the quasiparticle distribution function starting with the case of zero phonon temperature and then generalizing to finite temperature. Interestingly we will show that despite the assumption of low $T_B$, stimulated emission and absorption of phonons cannot be neglected at all energies and, in fact, determine the high-energy tail of the distribution function. 

\section{Shape of the quasiparticle distribution function}
\label{sec: Shape of the quasiparticle distribution}

As discussed in the Introduction, our goal is to find the quasiparticle distribution function by approximately solving the system of coupled equations Eqs.~(\ref{sec:Rate Equation eq: Rate Equation}), (\ref{sec:Rate Equation eq: Gap Equation}), and (\ref{sec:Rate Equation eq: Phonon Rate Equation}) in the steady state in the regime of low temperatures, and hence low quasiparticle density, and high number of photons, $\bar{n} \gg 1$, or equivalently $T_0 \gg \omega_0$. A sufficient condition for having low density is $f(E) \ll 1$, which enables us to approximate the Pauli-blocking factors as $[1-f(E)] \approx 1$. Moreover, we neglect in this section the recombination and pair-breaking collision integrals, Eqs.~\eqref{sec:Rate Equation eq: Gamma^Phon_re} and \eqref{sec:Rate Equation eq: Gamma^Phon_PB}, since they affect the normalization but not the shape of the quasiparticle distribution function, as argued in Sec.~\ref{sec:Lifetimes}.

At low temperature $T_B \ll \omega_0$, the competition between absorption of photons and emission of phonons results in a quasiparticle distribution with peaks at energies $\Delta+m\omega_0$, $m=0,\,1,\,2,\ldots$~\cite{Goldie.2012,G.Catelani.2019}. Here we focus on the envelope function that determines the heights of these peaks. Then interpreting $f(E)$ as this envelope, the photon collision integral Eq.~\eqref{sec:Rate Equation eq:  Photon integral} can be approximately written as a generalized diffusion operator in energy space,
\begin{align}\label{sec:Shape of the quasiparticle distribution eq: Diffusion Operator}
    &St^{Phot}\{f,\bar{n}\}\simeq c_{Phot}^{QP}\frac{\omega_{0}^2}{U^+(E,E)}\bar{n}\\
    &\times\frac{\partial}{\partial E'}\left[U^+(E,E')^2e^{-E'/T_0}\frac{\partial}{\partial E'}\left(f(E')e^{E'/T_0}\right) \right]\bigg|_{E'=E} \nonumber
\end{align}
as one can verify by Taylor expansion of Eq.~\eqref{sec:Rate Equation eq:  Photon integral} to second order in $\omega_0$. Next, we consider explicitly three cases:  A. phonons in equilibrium at zero temperature; B. phonons in equilibrium at finite temperature; C. nonequilibrium corrections to the phonon distribution function.

\subsection{Phonons in equilibrum at \texorpdfstring{$T_B=0$}{TB=0}}
\label{sec:zeroTphonons}

The assumption that phonon are in equilibrium corresponds to taking the limit $\tau_l \to 0$ in Eq.~(\ref{sec:Rate Equation eq: Phonon Rate Equation}), so that $n(\omega) = n_T(\omega,T_B)$ is the (leading order) solution for the phonon distribution function. For $T_B = 0$, this implies $n(\omega) = 0$, and the steady-state equation for the quasiparticle distribution function reduces to
\begin{equation}\label{sec:Shape of the quasiparticle distribution eq: Ansatz analytic}
0 = St^{Phon}_{sp}\{f\}+St^{Phot}\{f,\bar{n}\}
\end{equation}
with $St^{Phon}_{sp}$ of Eq.~(\ref{sec:Rate Equation eq: Gamma^Phon_sp}) and $St^{Phot}$ of Eq.~(\ref{sec:Shape of the quasiparticle distribution eq: Diffusion Operator}). Even this much simplified equation cannot be solved exactly, so we consider separately three energy ranges -- low, intermediate, and high -- to be defined below. In all three ranges we assume the photon number to be so large that $T_0$ is the highest energy scale; then we can take the limit $T_0 \to \infty$ in Eq.~(\ref{sec:Shape of the quasiparticle distribution eq: Diffusion Operator}) and replace the exponential factors with unity.

The low and intermediate energy ranges are sufficiently close to the gap, $E-\Delta \ll \Delta$, so that
one can approximate 
\begin{equation}\label{eq:Uplowen}
  U^+(E_1,E_2)\simeq \sqrt{\frac{2\Delta}{E_2-\Delta}}  
\end{equation} 
for $E_2>\Delta$, and $U_+ = 0$ otherwise, in Eq.~(\ref{sec:Shape of the quasiparticle distribution eq: Diffusion Operator}) and 
\begin{equation}\label{sec:Shape of the quasiparticle distribution eq: Low Energy approximation U_-}
    U^-(E_1,E_2)\simeq \frac{E_1+E_2-2\Delta}{\sqrt{2\Delta(E_2-\Delta)}}
\end{equation} in Eq.~(\ref{sec:Rate Equation eq: Gamma^Phon_sp}). Within this approximation, $St^{Phon}_{sp}\{f\}$ takes the form
\begin{align}\label{sec:Shape of the quasiparticle distribution eq: Low Energy approximation Phonon term}
    St^{Phon}_{sp}\{f\}&=-\frac{128}{105\sqrt{2}}\left(\frac{\Delta}{T_c}\right)^3\left(\frac{E-\Delta}{\Delta}\right)^{7/2}\frac{f(E)}{\tau_0}\\
    &+\frac{1}{\tau_0   T_c^3} \int\limits_{0}^{\infty} \!d\omega\,\omega^2 \frac{2E-2\Delta+\omega}{\sqrt{2\Delta (E+\omega-\Delta)}} f(E+\omega) \nonumber
\end{align}
Introducing the temperature scale 
\begin{equation}\label{eq:Tstardef}
    T_* \equiv \left( \frac{105}{64} T_c^3 c^{QP}_{Phot}\bar{n}\tau_0
    \omega_0^2 \Delta \right)^{1/6}
\end{equation}
characterizing the width of the distribution function, and neglecting the first term on the right-hand side in Eq.~\eqref{sec:Shape of the quasiparticle distribution eq: Low Energy approximation Phonon term} for $E-\Delta\lesssim T_*$ (low energy range) and the second one for $T_* \lesssim E-\Delta \lesssim \Delta$ (intermediate range) leads to the solution derived in Ref.~\cite{G.Catelani.2019} 
\begin{numcases}{f(x)\simeq}
 b_0 \left(1-0.564 x^{5/2}+0.119 x^{7/2}\right) , \; x\lesssim 1 \quad \label{sec: Shape of the quasiparticle distribution eq: Low Energy Solution low energy}\\
 3 b_0 \mathrm{Ai}\left(\frac{x^2}{4^{1/3}}\right) , \; 1 \lesssim x\ll \Delta/T_* \label{sec: Shape of the quasiparticle distribution eq: Low Energy Solution high energy}
\end{numcases}
with $x\equiv (E-\Delta)/T_*$, $\mathrm{Ai}$ the Airy function, and $b_0$ a normalization constant whose determination is the subject of Sec.~\ref{sec: Quasiparticle Density}. Note that depending on the parameters (and in particular by increasing $\bar{n}$) the intermediate regime could be absent; here we assume for simplicity that the condition $T_* < \Delta$ is satisfied (this is consistent with considering quasiparticles with energies up to few times the gap, cf. Sec.~\ref{sec:Lifetimes}). Moreover, the initial assumptions that $T_0$ is large and that we can study the envelope concretely means $T_0 \gg T_* \gg \omega_0$.

In the high-energy range $E-\Delta\gtrsim \Delta$, we neglect terms of order $(\Delta/E)^2$ in Eq.~\eqref{sec:Shape of the quasiparticle distribution eq: Diffusion Operator}, leading to
\begin{equation}\label{sec: Shape of the quasiparticle distribution eq: Approx Gamma^Phot}
    St^{Phot}\{f,\bar{n}\}= \bar{n}c^{QP}_{Phot}\omega_{0}^2 f''(E)
\end{equation}
Since for $T_*<\Delta$ most quasiparticles are at energies below $2\Delta$, at energies $E>2\Delta$ the first integral in Eq.~\eqref{sec:Rate Equation eq: Gamma^Phon_sp} is much smaller than the second one and can be neglected. Therefore we can further approximate
\begin{equation}\label{eq:ApproxGamma^Phon_sp}
St^{Phon}_{sp}\simeq -\frac{f(E)}{\tau_0 T_c^3}\frac{(E-\Delta)^3}{3}\
\end{equation}
in the high-energy regime.
Substituting $\tilde{x}\equiv(E-\Delta)/\tilde{T}_*$, with
\begin{equation}\label{eq:tildeTstar_def}
    \tilde{T}_* \equiv(3\bar{n}c^{QP}_{Phot}\omega_0^2 \tau_0 T_c^3)^{1/5} =\left(\frac{64}{35}\frac{T_*}{\Delta}\right)^{1/5}T_* \,,
\end{equation}
into Eq.~\eqref{sec:Shape of the quasiparticle distribution eq: Ansatz analytic}, that equations takes the form of a generalized Airy equation,
\begin{equation}\label{sec: Shape of the quasiparticle distribution eq: A_3 eq}
    f''(\tilde{x})-\tilde{x}^3f(\tilde{x})=0 \, .
\end{equation}
The solution to this equation can be written in terms of a modified Bessel function of the second kind,
\begin{equation}\label{sec: Shape of the quasiparticle distribution eq: High Energy Solution}
f(\tilde{x})=\tilde{b}_0 \sqrt{\tilde{x}} K_{1/5}\left(\frac{2}{5}\tilde{x}^{5/2}\right)    
\end{equation}
With our assumptions we are only interested in the limit of large $\tilde{x}$, so that we can approximate
\begin{equation}
\label{sec: Shape of the quasiparticle distribution eq: High Energy Solution approximate}
    f(\tilde{x}) \simeq \tilde{b}_0 \sqrt{\frac{5\pi}{4\tilde{x}^{3/2}}}e^{-2\tilde{x}^{5/2}/5}
\end{equation}
This expression should match the similar approximation for Eq.~(\ref{sec: Shape of the quasiparticle distribution eq: Low Energy Solution high energy}) at an energy of order $2\Delta$. Indeed, the exponential factors are identical at $E=143\Delta/80$ and the prefactors are then related by
\begin{equation}
    \frac{\tilde{b}_0}{b_0} = \frac{3^{3/2}2^{2/3}}{5\pi}\left(\frac{64}{35}\frac{T_*}{\Delta}\right)^{-2/5}
\end{equation}
Due to the faster than exponential decay of the distribution over the energy scale $T_* \lesssim \Delta$, it might seem irrelevant to calculate here and in the next subsection the behavior of its high-energy tail ($E \gtrsim 2\Delta$); however, quasiparticles with energy above $3\Delta$ can relax by emitting a pair-breaking phonon, which in turn can generate two quasiparticles. That is why knowledge of the tail will be needed to understand how the photons influence the quasiparticle density, see Sec.~\ref{sec: Quasiparticle Density}.

\subsection{Equilibrium phonons, \texorpdfstring{$T_B>0$}{TB>0}}
\label{sec:finteTphonons}

We now consider in more detail the effect of thermal phonons. We aim to show that neglecting phonons is a good approximation up to a crossover energy $E_*$, above which the (envelope of the) quasiparticle distribution function takes the thermal equilibrium form. Using the expression in Eq.~\eqref{sec:Rate Equation eq: Stimulated Phonon integral},
we define the energy scale $E_*$ as that energy at which the equation
 \begin{equation}\label{sec: Shape of the quasiparticle distribution eq: Definition E_*}
    St^{Phon}_{st}\{f,n_T\}+St^{Phot}\{f,\bar{n}\} = 0
\end{equation}
is satisfied with $f$ as obtained in Sec.~\ref{sec:zeroTphonons} and $n= n_T$ being the thermal distribution for phonons.

By the definition of $E_*$, up to that energy we can neglect the phonons and hence Eq.~\eqref{sec:Shape of the quasiparticle distribution eq: Ansatz analytic} holds; that equation enables us to express $St^{Phot}$ in terms of $St^{Phon}_{sp}$. For the latter, at high energies $E -\Delta \gg \Delta$ we can use Eq.\eqref{eq:ApproxGamma^Phon_sp} to get:
\begin{equation}\label{eq:StPhotHighE}
    St^{Phot}\{f,\bar{n}\}=-St^{Phon}_{sp}\{f\} \simeq \frac{(E-\Delta)^3}{3\tau_0 T_c^3}f(E)
\end{equation}
with $f(E)$ decreasing faster than exponentially, cf. Eq.~\eqref{sec: Shape of the quasiparticle distribution eq: High Energy Solution approximate}. To estimate $E_*$ using Eq.~\eqref{sec: Shape of the quasiparticle distribution eq: Definition E_*} we need an approximate expression for $St_{st}^{Phon}$. It turns out that the main contribution to this collision integral originates from the term in Eq.~\eqref{sec:Rate Equation eq: Stimulated Phonon integral} proportional to $f(E-\omega)$, since as function of $\omega$ that factor increases faster than exponentially (as long as $E-\omega$ remains sufficiently large). Thus it dominates over the exponential suppression of $n(\omega,T_B)$ at $\omega\gg T_B$, leading to a sharply peaked maximum of the integrand at a certain energy $\omega_M$,
as detailed in Appendix~\ref{appendix: Laplace approximation}. 
Introducing the crossover temperature
\begin{equation}\label{eq:TBstardef}
    T_B^* \equiv \left(\frac{T_*}{\Delta}\right)^3 \Delta = \omega_0 \sqrt{\frac{105}{64}c^{QP}_{Phot} \bar{n}\tau_0} \left(\frac{T_c}{\Delta}\right)^{3/2}
\end{equation}
we can distinguish two regimes: for low phonon temperature/high photon number, $T_B \ll T_B^*$, we find 
\begin{equation}\label{sec: Shape of the quasiparticle distribution eq: Approximate E_* High E}
E_* \approx \Delta + \tilde{T}_*(\tilde{T}_*/T_B)^{2/3} \gtrsim 2\Delta \, ,
\end{equation}
while for high phonon temperature/low photon number, $T_B^* \lesssim T_B \ll T_*$, we get \begin{equation}\label{sec: Shape of the quasiparticle distribution eq: Approximate E_* Low E}
     E_* \approx \Delta + T_* (T_*/T_B)^{1/2} \lesssim 2\Delta \, .
\end{equation}
In both cases, it turns out that for $E>E_*$ stimulated emission/absorption of phonons dominates over the interaction with photons, so that the distribution function is approximately of the Boltzmann form,
\begin{equation}\label{eq:f_boltzmann}
f(E) \simeq b_T e^{-E/T_B},
\end{equation}
where $b_T$ can be found by requiring continuity of $f$ at $E=E_*$.
Note that while here the ratio between $T_B$ and $T_B^*$ being below or above unity has the apparently minor role of determining whether $E_*$ is above or below $2\Delta$, we will later see that this ratio influences also the temperature dependence of both the quasiparticle density and the quality factor.

\subsection{Finite thermalization time}
\label{sec:Shape of the quasiparticle distribution Phonon trapping}

So far we have assumed that the phonon distribution has the equilibrium form, corresponding to the limit of zero thermalization time $\tau_l$. If the thermalization time is non-zero, 
the phonon distribution can deviate from the equilibrium one, as discussed in Sec.~\ref{sec:Lifetimes}. We can distinguish between phonons of energy $\omega$ below and above the pair-breaking threshold $2\Delta$. Above-threshold phonons can break Cooper pairs and thus influence the quasiparticle density -- this is the subject of Sec.~\ref{sec: Quasiparticle Density}. These phonons can also affect the shape of the quasiparticle distribution by being absorbed, but these processes are far less frequent than pair breaking (since $\tau^{Phon}_{abs}\gg \tau_0^{PB}$) and the change would only take place at high energies $E>3\Delta$ where the occupation is generically extremely small, so we neglect this effect.

Below-threshold phonons, in contrast, can affect the quasiparticle distribution at all energies. However, we now show that significant deviations from the equilibrium phonon distribution appear only at relatively high energy $\omega$ and have a negligible effect on the shape of the quasiparticle distribution. Indeed, for $\omega<2\Delta$ we can approximately solve Eq.~\eqref{sec:Rate Equation eq: Phonon Rate Equation} in the steady state, writing $n(\omega) \simeq n_T(\omega,T_B) + n_1(\omega)$ with (see Appendix~\ref{appendix:noneqphonons} for details)
\begin{equation}
\label{sec:Shape of the quasiparticle distribution eq: Definition n_1}
    n_1(\omega)\simeq\frac{2 \tau_l}{\pi \Delta_0 \tau_0^{PB}}\!\int\limits_{\Delta}^{\infty}\!dE \, \rho(E) U^-(E,E+\omega) f(E+\omega) 
\end{equation}
and the $n_1$ term becomes dominant above the crossover energy
\begin{equation}\label{sec:Shape of the quasiparticle distribution eq: omegac}
    \omega_c \approx T_B \ln \left(\frac{\Delta}{T_*} \frac{\tau_0^{PB}}{\tau_l} b_0^{-1}\right)
\end{equation}
(this approximate expression is valid for $T_* \lesssim \omega_c \ll T_*\sqrt{3T_*/T_B}$ and $\omega_c\lesssim \Delta$, see Appendix~\ref{appendix:noneqphonons} for a more accurate determination of $\omega_c$). The crossover energy depends on the quasiparticle density through $b_0\ll 1$, which cannot be determined without considering the non-linear terms in the kinetic equations. However, for thermalization time $\tau_l$ short compared to $\tau_0^{PB}$, we expect the density to be comparable to the one in thermal equilibrium and hence $b_0 \sim e^{-\Delta/T_B}$ (the quasiparticle density is discussed more in detail in Sec.~\ref{sec: Quasiparticle Density}); this implies that $\omega_c$ would become larger than $\Delta$ and grow with decreasing $\tau_l$ or $T_*$, as one would expect, since reducing these parameters means that the system is closer to thermal equilibrium. 
We will return to this point when comparing our results to numerical calculations in Sec.~\ref{sec:NumComp}.

To determine if $n_1$ influences the shape of the distribution function $f$, one can proceed as in Sec.~\ref{sec:finteTphonons}, by replacing $n_T \to n_T+n_1$ in Eq.~\eqref{sec: Shape of the quasiparticle distribution eq: Definition E_*}. As mentioned there, the main $n_T$ contribution to the collision integral $St_{st}^{Phon}$ comes from the region around an energy $\omega_M$; if this energy is smaller than $\omega_c$, we expect negligible impact of $n_1$ on the shape of $f$.
This is clearly the case in the limit of fast phonon thermalization $\tau_l \to 0$, since in this case, as discussed above, $\omega_c> \Delta$, while in general $\omega_M < \Delta$ (see Appendix~\ref{appendix: Laplace approximation}), so $\omega_M < \omega_c$.
We do not investigate here more generally when the condition $\omega_M < \omega_c$ is satisfied, nor the effect of $n_1$ on $f$ when it is violated, but we will show numerically that for experimentally relevant parameters the shape of $f$ derived in this section is valid at least up to energies of a few times $\Delta$. 

\section{Quasiparticle Density}
\label{sec: Quasiparticle Density}

The considerations in the previous section have been limited to the shape of the quasiparticle distribution function, including the effect of nonequilibrium phonons due to finite thermalization time, see Sec.~\ref{sec:Shape of the quasiparticle distribution Phonon trapping}. However, it follows from the pair-breaking phonon collision integral, Eq.~\eqref{sec:Rate Equation eq: Gamma^Phon_PB}, that nonequilibrium phonons can potentially  affect the quasiparticle number as soon as the crossover frequency $\omega_c$ between equilibrium and nonequilibrium phonon population satisfies $\omega_c \lesssim 2\Delta$, a weaker condition than that required for them to affect the shape of the distribution function, $\omega_c < \omega_M$. Consequently in this section we investigate the effect a non-zero thermalization time has on the number of quasiparticles in the regime $\omega_M<\omega_c\lesssim 2\Delta$, in which the influence of nonequilibrium phonons on the quasiparticle distribution shape can be neglected while their influence on the quasiparticle density must be established. 

The quasiparticle density $N_\mathrm{qp}$ is given by
\begin{equation}\label{eq:Nqp}
    N_\mathrm{qp} \equiv 4\rho_F \int_\Delta^\infty dE \, \rho(E) f(E) \simeq 4.2\rho_F \sqrt{2\Delta T_*} b_0 
\end{equation}
where for $f$ we used the distribution function of Sec.~\ref{sec: Shape of the quasiparticle distribution} and the numerical prefactor was determined in Ref.~\cite{G.Catelani.2019}. As remarked previously, to find the value of the normalization constant $b_0$, the recombination and pair-breaking collision integrals, Eqs.~\eqref{sec:Rate Equation eq: Gamma^Phon_re} and \eqref{sec:Rate Equation eq: Gamma^Phon_PB}, must be taken into account. To do so, we multiply the kinetic equation Eq.~\eqref{sec:Rate Equation eq: Rate Equation} by the BSC density of states $\rho(E)$, integrate over energy $E$, and assume the steady-state condition $df/dt=0$ to arrive at the equation
\begin{equation}
\begin{split}
    \int_{\Delta} dE \, \rho(E) \int_{E+\Delta}d\omega \, \omega^2 U^+(E,\omega-E) \\ \times\left[n(\omega)-f(E)f(\omega-E)\right] = 0
\end{split}\end{equation}
where, as before, we neglect Pauli-blocking factors and assume $n(\omega) \ll 1$ for $\omega > 2\Delta$. The two terms in square brackets originate from pair breaking and recombination, respectively. For the former, we can switch the integration order and realize that the resulting integral over $E$ is the same as that determining the lifetime of phonons against pair breaking (cf. Appendix~\ref{appendix:Lifetimes}); neglecting again the weak dependence of the result on $\omega$, we rewrite the above equation as
\begin{align} \label{eq:normalization}
    & \int_{2\Delta} d\omega \, \omega^2 \Bigg[n(\omega) - \\ & \frac{1}{\pi\Delta}\int_{\Delta}^{\omega-\Delta} dE \, \rho(E) U^+(E,\omega-E) f(E)f(\omega-E)
     \Bigg] = 0 \nonumber
\end{align}
To proceed further, we need to know the phonon distribution function above the pair-breaking threshold, $\omega > 2\Delta$; as shown in Appendix~\ref{appendix:noneqphonons}, it takes the form
\begin{equation}\label{eq:nnoneq2D}
    n(\omega)\simeq \frac{\tau_0^{PB}}{\tau_l + \tau_0^{PB}} \left[n_T(\omega,T_B) + n_1(\omega) +n_2(\omega) \right]
\end{equation}
with $n_1$ of Eq.~\eqref{sec:Shape of the quasiparticle distribution eq: Definition n_1} and $n_2$ being equal to the product of $\tau_l/\tau_0^{PB}$ times the second term in square bracket in Eq.~\eqref{eq:normalization}.

Equation \eqref{eq:normalization} can now be recast as a quadratic equation for $b_0$ (see Appendix~\ref{appendix: Neglected terms in the Normalization})
\begin{equation}\label{eq:Normalization eq}
    I_2 b_0^2 - 2\frac{\tau_l}{\tau_0^{PB}}I_1 b_0 - I_0 = 0
\end{equation}
with the quadratic term arising from the pair-breaking contribution together with $n_2$, the linear term from $n_1$, and the constant term from $n_T$. Due to their origins, $I_0$ and $I_2$ are dimensionless functions of $T_B/\Delta$ and $T_*/\Delta$ respectively, while $I_1$ depends in general on both. We can distinguish two limiting cases: if $(\tau_l I_1/\tau_0^{PB})^2 \ll I_2 I_0$, then (cf. Appendix~\ref{appendix: Neglected terms in the Normalization})
\begin{equation}\label{eq:b0TB}
    b_0 \simeq \sqrt{I_0/I_2} \simeq \sqrt{\pi T_B/T_*}e^{-\Delta/T_B}/2.1
\end{equation}  
and the quasiparticle density is approximately the same as in thermal equilibrium, even though the distribution function differs significantly from the equilibrium one.
In the opposite limit we find that $b_0 \simeq 2\tau_l I_1/\tau_0^{PB} I_2$ is larger than the thermal equilibrium value and is proportional to the ratio $\tau_l/\tau_0^{PB}$: the larger this ratio (that is, the longer $\tau_l$), the easier it is to enter into this limit and the larger the quasiparticle density.
We discuss further the corresponding nonequilibrium quasiparticle density in the framework of a generalized Rothwarf-Taylor model.

\subsection{Generalized RT model}
\label{sec:genRT}

Given the proportionality between $b_0$ and $N_\mathrm{qp}$, Eq.~\eqref{eq:Nqp}, the last and first terms in the left-hand side of Eq.~\eqref{eq:Normalization eq} correspond exactly to the phonon generation and quasiparticle recombination terms in the steady-state version of the Rothwarf-Taylor model~\cite{Rothwarf.1967} (we do not consider here direct quasiparticle injection). In fact, we can relax the steady-state assumption and allow for variation in time of parameters such as bath temperature ($T_B$) and photon number (\textit{i.e.}, $T_*$), as long as their change is slow on the scale over which the shape of the distribution function is established, namely the quasiparticle scattering times, see Sec.~\ref{sec:Lifetimes} (note that since $\tau_0^{PB}$ is shorter that $\tau_0$, the phonon distribution quickly follows any change in the quasiparticle one). Then the shape of the distribution function is at all times the one we have calculated in Sec.~\ref{sec: Shape of the quasiparticle distribution} and integration over energy of the kinetic equation times $4\rho_F \rho(E)$ gives
\begin{equation}\label{eq:genRT}
    \frac{dN_\mathrm{qp}}{dt} = G_T + G(T_*/\Delta) N_\mathrm{qp} - R N_\mathrm{qp}^2
\end{equation}
Here
\begin{equation}
    G_T = \frac{16\pi \rho_F\Delta}{\bar{\tau}_0}\left(\frac{\Delta}{T_c}\right)^3  \frac{T_B}{\Delta} e^{-2\Delta/T_B}
\end{equation}
is the rate of quasiparticle generation (per unit volume) due to thermal phonons for $T_B$ small compared to $2\Delta$, and 
\begin{equation}\label{eq:Rdef}
    R = \frac{2\Delta^2}{\rho_F \bar{\tau}_0 T_c^3} 
\end{equation}
is the quasiparticle recombination coefficient. Both quantities are renormalized by the finite phonon thermalization time, an effect known as ``phonon trapping''\cite{Chang.1978},
\begin{equation}\label{eq:tau0bar}
    \bar{\tau}_0 = \tau_0 (1+ \tau_l/\tau_0^{PB})
\end{equation}
Note that as $T_*$ approaches $\Delta$, corrections to $R$ and $\tau_0^{PB}$ resulting from the finite distribution's width $\sim T_*/\Delta$ can become relevant, as discussed in Appendix~\ref{appendix: Higher Order Corrections to the Quasiparticle density}.

\begin{figure}
    \includegraphics[scale=0.5]{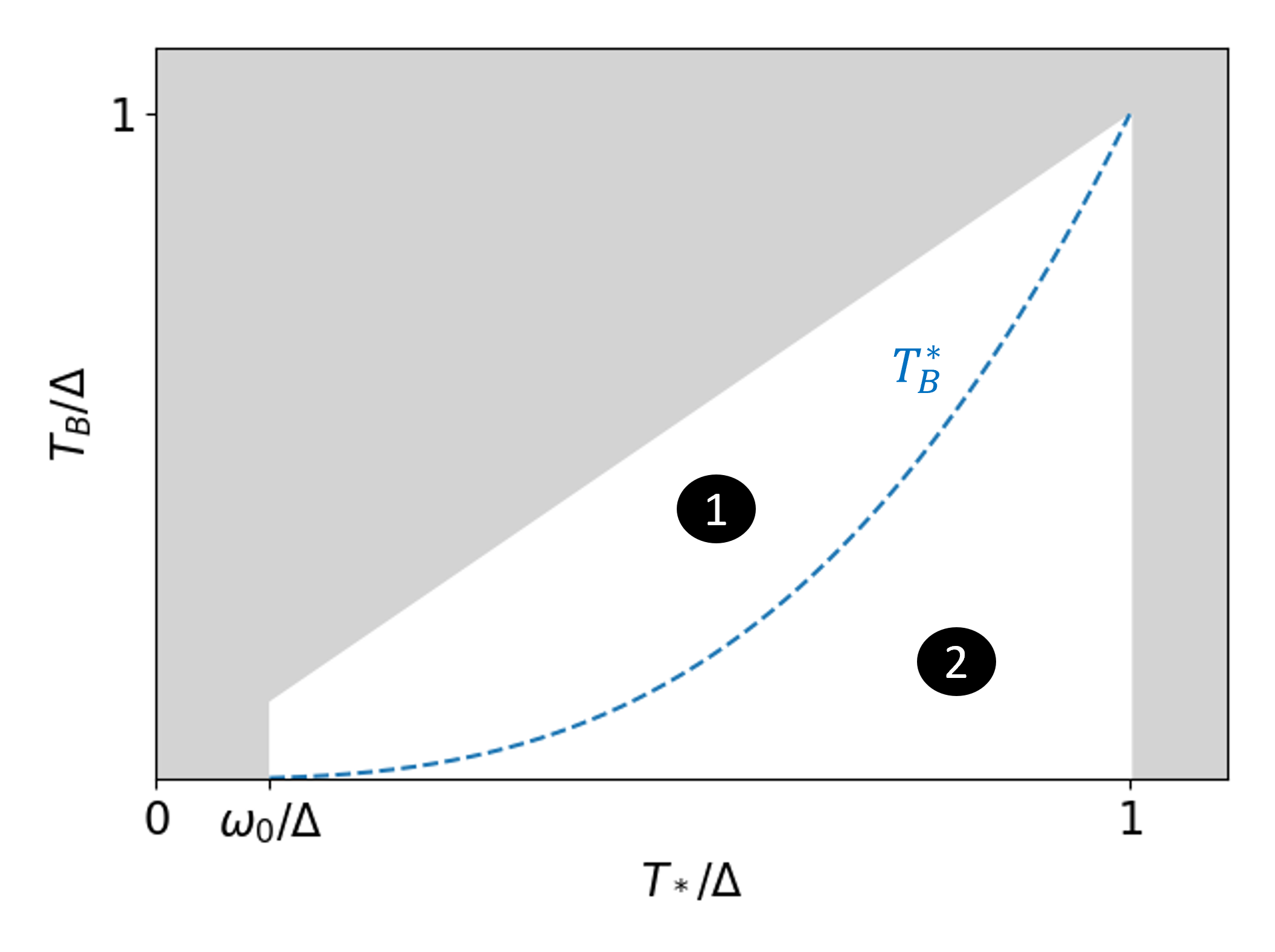}
    \caption{Depending on the phonon temperature $T_B$ and the photon number $\bar{n}$ [that is, $T_*$ of Eq.~\eqref{eq:Tstardef}], there are two regimes for the quasiparticle density: in regime 1, thermal phonons dominate quasiparticle creation and the quasiparticle density is approximately as in equilibrium. In regime 2, the photons drive a sufficient amount of quasiparticles to energies $E>3\Delta$ such that quasiparticle creation from phonons emitted by these high-energy quasiparticles is dominant; in this regime, the quasiparticle density is larger than in equilibrium and depends on the photon number but not on the temperature of the phonon bath. The crossover between the two regimes takes places approximately when $T_B\sim T_*^3/\Delta^2$ [cf. Eq.~\eqref{eq:TBstardef}], as indicated by the dashed curve. The area shaded in gray ($T_*< \omega_0$, $T_*>\Delta$, $T_B>T_*$) identifies the parameter regions where approximations made in the analytical derivations are not valid.}
\label{fig: Different Regimes Overview}
\end{figure}

The central term in the righ-hand side of Eq.~\eqref{eq:genRT} is absent in the Rothwarf-Taylor model and represent an additional quasiparticle generation term proportional to the quasiparticle density itself. It originates from pair-breaking nonequilibrium phonons emitted by quasiparticles which have been excited to sufficiently high energies by the photons. Indeed, the coefficient can be taken in the form (see end of Appendix~\ref{appendix: Neglected terms in the Normalization})
\begin{equation}\label{eq:Gdef}
    G(x) = \frac{\gamma}{\bar{\tau}_0} \frac{\tau_l}{\tau_0^{PB}} \left(\frac{\Delta}{T_c}\right)^3 \, x^{9/2} e^{-\sqrt{14/5}x^{-3}}
\end{equation}
with $\gamma = 2^{13/6}3^{3/2}/(2.1\times5 \sqrt{7}) \simeq 0.84$, and vanishes if the phonons are forced to be in thermal equilibrium ($\tau_l=0$). At high phonon temperature/low photon number, $T_B \gtrsim T_B^*$, this term can be neglected and in the steady state the quasiparticle density takes the thermal equilibrium value; in the opposite regime, $T_B \ll T_B^*$, it leads to a quasiparticle density independent of the bath temperature $T_B$ and larger than in thermal equilibrium.  Although the linearity in $N_\mathrm{qp}$ of this additional generation term could be expected -- the more quasiparticle there are, the more can be excited to high energy and emit phonons -- we stress that the dependence on the photon number can be found only after solving the kinetic equation for the shape of the distribution function; therefore, it is beyond the reach of phenomenological treatments that consider just the quasiparticle density from the outset. In fact, in this regime the quasiparticle density is strongly dependent on the photon number; the strong dependence originates from the fact that only quasiparticles in the high-energy tail of the distribution function, $E>3\Delta$, can emit pair-breaking photons, and a quasiparticle must absorb a large number of photons to reach that energy while also losing energy by emitting phonons. Note that in the extreme case $T_B=0$, the solution $N_\mathrm{qp}$=0 is unstable: even a single quasiparticle can start the process of driving the phonons out of equilibrium and hence generate more quasiparticles.
In Fig.~\ref{fig: Different Regimes Overview} we provide an overview of the different regimes we have identified and of the parameter regions where our approach is applicable.
\subsection{Gap suppression}

Once the normalization constant $b_0$ is found using Eq.~\eqref{eq:Normalization eq} [or, equivalently, $N_\mathrm{qp}$ using Eq.~\eqref{eq:genRT}], we can perturbatively calculate the change in the gap $\delta\Delta=\Delta_0 - \Delta$. Indeed, substituting the distribution function of Sec.~\ref{sec: Shape of the quasiparticle distribution} into  Eq.~\eqref{sec:Rate Equation eq: Gap Equation} we find
\begin{equation}\label{eq:gapsupp}
    \frac{\delta\Delta}{\Delta_0} = 4.2 b_0 \sqrt{\frac{T_*}{2\Delta_0}} \left(1-\frac14 \frac{0.88}{2.1}\frac{T_*}{\Delta_0}\right)
\end{equation}
where we used the numerical estimate $\int_0 dx \sqrt{x} f(x) \simeq 0.88$ (in the perturbative calculation we neglect the deviation of $\Delta$ from $\Delta_0$ in the right-hand side of Eq.~\eqref{sec:Rate Equation eq: Gap Equation}; this is consistent if $T_*/\Delta_0 \gg b_0^2$). The leading order term is equal to the leading-order approximation for $N_\mathrm{qp}/(2\rho_F \Delta_0)$ [cf. Eq.~\eqref{eq:Nqp}], which is the fraction of broken Cooper pairs; this result is generic to quasiparticles whose distribution function width above the gap (in our case, $T_*$) is small compared to the gap itself. 
Including corrections up to second order in $T_*/\Delta$, we can rewrite Eq.~\eqref{eq:gapsupp} in the form
\begin{equation}\label{eq:gapsupp_N}
    \frac{\delta\Delta}{\Delta_0} = \frac{N_\mathrm{qp}}{2\rho_F \Delta_0} \left[1-0.42\frac{T_*}{\Delta_0}+0.22\left(\frac{T_*}{\Delta}\right)^2\right]
\end{equation}
For comparison, in thermal equilibrium the terms in brackets read $1-0.5 T_B/\Delta_0+3(T_B/\Delta_0)^2/8$; this shows that for a given quasiparticle density, the gap is less suppressed in the nonequilibrium case, since by assumption $T_* > T_B$. We can therefore find an enhancement of superconductivity, since for $T_B \gtrsim T_B^*$ the quasiparticle density takes roughly the same value as in equilibrium (more accurately, the quasiparticle density slightly decreases with increasing $T_*$ in this regime, as discussed in Appendix~\ref{appendix: Higher Order Corrections to the Quasiparticle density}, and thus the enhancement is even stronger). In the opposite case, the density is much larger than in equilibrium and hence superconductivity is weakened. 

\subsection{Comparison to numerics}
\label{sec:NumComp}

As a validation of our approach, we now compare the analytical results to the numerical solution of the full system of kinetic equations, Eqs.~\eqref{sec:Rate Equation eq: Rate Equation}-\eqref{sec:Rate Equation eq: Phonon Rate Equation}. Unless otherwise stated,in this subsection we use the parameters listed in Table~\ref{tab: Param 1}. Their values have been chosen to enable the comparison to experiments that will be discussed in the next section; that is, they should be typical for thin aluminum film resonators. The critical temperature is assumed to be connected to the gap via the BCS relation $\Delta_0=1.764 T_c$, resulting in $T_c\simeq 1.18\,$K. Using Eq.~\eqref{sec:Rate Equation eq: Relation Lifetimes} and the assumed $\tau_0$, the phonon lifetime against pair breaking is $\tau_0^{PB}=255\,$ps. The parameters loosely satisfy the validity conditions for the analytical approximations, namely $\omega_0\ll T_* \lesssim \Delta$, $b_0\ll 1$, and $\omega_c > \omega_M$ ($b_0$ can be read off Fig.~\ref{fig:Distributionf}; using the results of Appendix~\ref{appendix: Laplace approximation} we calculate $\omega_M \simeq 0.57 \Delta_0$, while $\omega_c$ can be estimated from Fig.~\ref{fig:Distributionn}). For the numerical calculations, we take $h=\Delta_0/180$ for the discretization step size and truncate the energy at $E_\mathrm{max} = 10\Delta_0$; note that $\omega_0 = 20h$, meaning the shape of the peaks is captured by the numerics.

\begin{table}[!tb]
\centering
\begin{tabular}{|c|c|c|c|c|c|c|c|c|c|}
\hline
$c^{QP}_{Phot}$&$\tau_0$ & $\Delta_0$ &$\omega_0$ & $\bar{n}$& $T_B$  & $T_*$& $T_B^*$\\
\hline
$1\,$Hz & $438\,$ns & $180\,\mu$eV & $\Delta_0/9$ & $10^7$ & $0.1\,$K & $\Delta_0/2\,$ & 0.26 K  \\
\hline
    \end{tabular}
    \caption{Parameter used for the plots in Figs.~\ref{fig:Distributionf} and \ref{fig:Distributionn}. The quantities $T_*$ and $T_B^*$ are calculated from the other parameters using their respective definitions, Eqs.~\eqref{eq:Tstardef} and \eqref{eq:TBstardef}.}
    \label{tab: Param 1}
\end{table}

In Fig.~\ref{fig:Distributionf} we plot with solid lines the numerically calculated quasiparticle distribution function $f$ as function of energy for different values of the thermalization time $\tau_l$. In all cases, we find good agreement with the analytical predictions (dashed lines) of Sec.~\ref{sec: Shape of the quasiparticle distribution} spanning several orders of magnitude in occupation probability, whose large variation takes place over an energy range of a few times the gap. For $\tau_l=0$, the phonons are at thermal equilibrium and the high-energy tail of $f$ approaches the expected exponential decay, see Sec.~\ref{sec:finteTphonons}. As $\tau_l$ increases, however, the high-energy tail deviates significantly from the $\tau_l=0$ prediction; the reason for this deviation is the re-absorption of phonons emitted by recombination processes [cf. $n_2$ in Eq.~\eqref{eq:nnoneq2D}], which have energy $\omega> 2\Delta$.  
Consequently, the deviation takes place at energies above $3\Delta$. At those energies for $T_* \lesssim \Delta$ the occupation probability is so small that the deviation does not affect the quasiparticle density. 

\begin{figure}
    \includegraphics[scale=0.5]{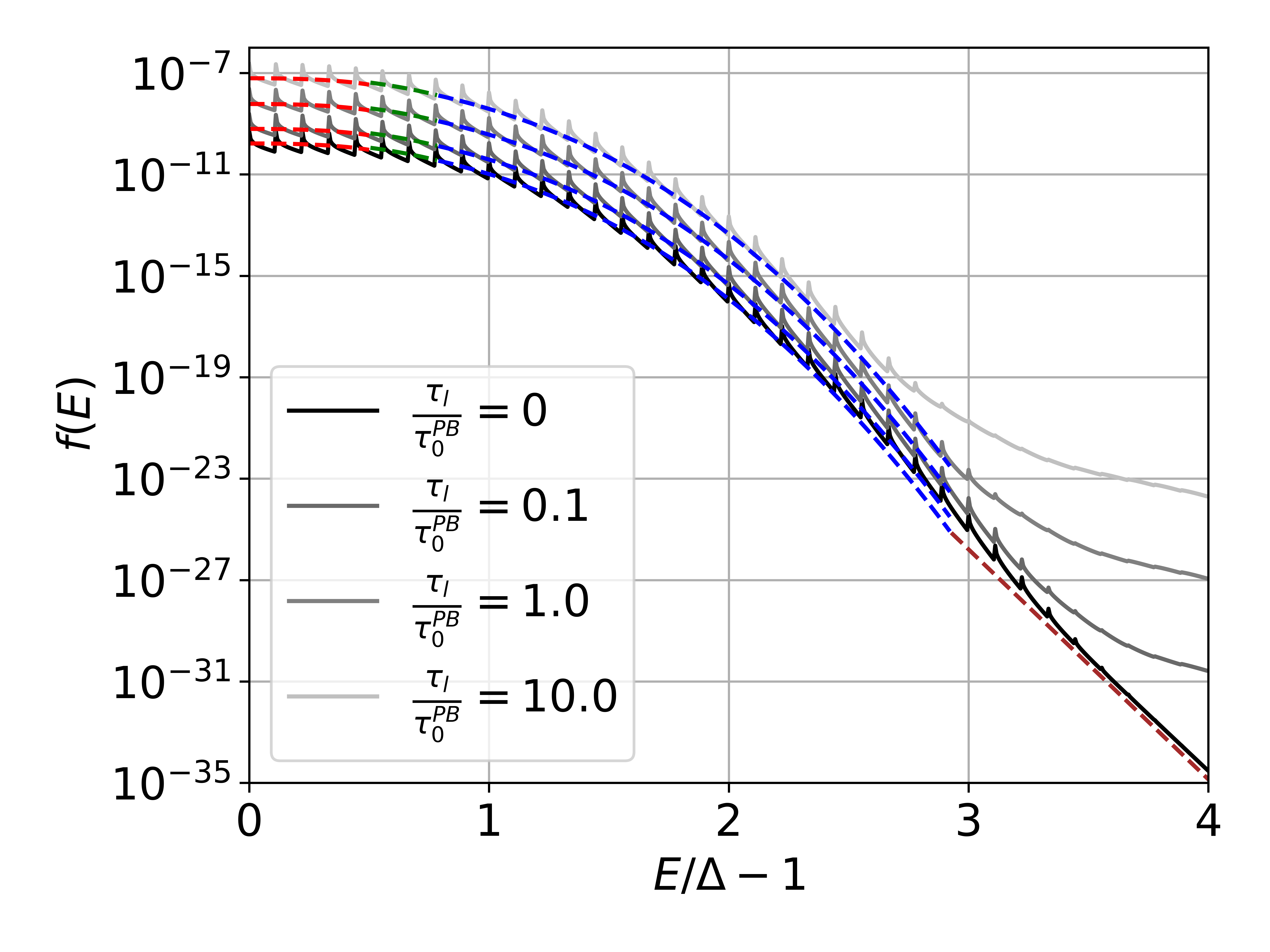}
    \caption{Quasiparticle distribution function vs energy for the parameters in Table~\ref{tab: Param 1}. Grey solid lines shows the results of numerical calculations. The colored dashed lines are obtained using the analytical approximations in their respective regimes of applicability; from left to right: Eq.~\eqref{sec: Shape of the quasiparticle distribution eq: Low Energy Solution low energy}, Eq.~\eqref{sec: Shape of the quasiparticle distribution eq: Low Energy Solution high energy}, Eq.~\eqref{sec: Shape of the quasiparticle distribution eq: High Energy Solution approximate}, and (for $\tau_l=0$) Eq.~\eqref{eq:f_boltzmann}.
    The normalization coefficient $b_0$ is calculated by using the steady-state solution to Eq.~\eqref{eq:genRT} in Eq.~\eqref{eq:Nqp} (we have taken into account the corrections to recombination coefficient and quasiparticle density discussed in Appendix~\ref{appendix: Higher Order Corrections to the Quasiparticle density}).}
    \label{fig:Distributionf}
\end{figure}

Figure~\ref{fig:Distributionn} shows the phonon distribution function versus energy $\omega$. There are no visible deviations from the thermal equilibrium behavior up to the $\tau_l$-dependent energy $\omega_c$, see Sec.~\ref{sec:Shape of the quasiparticle distribution Phonon trapping}, for all the values of $\tau_l$ considered. For $\omega$ between $\omega_c$ and $2\Delta$ phonons spontaneously emitted by the non-equilibrium quasiparticles become relevant and the phonon distribution function is predominantly given by $n_1(\omega)$ of Eq.~\eqref{sec:Shape of the quasiparticle distribution eq: Definition n_1} (see also Appendix~\ref{appendix:noneqphonons}).
For $\omega > 2\Delta$ the recombination of non-equilibrium quasiparticles affects the phonon distribution. Since the quasiparticle density is larger than in equilibrium, this leads to a phonon occupation probability bigger than the equilibrium one, as captured by the term $n_2(\omega)$ in Eq.~\eqref{eq:nnoneq2D} (see also Appendix~\ref{appendix:noneqphonons}, where we give an expression for $n_2$ valid up to $\omega\simeq2\Delta+T_*$; the analytically calculated phonon distribution is in good agreement with the numerical results).

\begin{figure}
    \centering
    \includegraphics[scale=0.5]{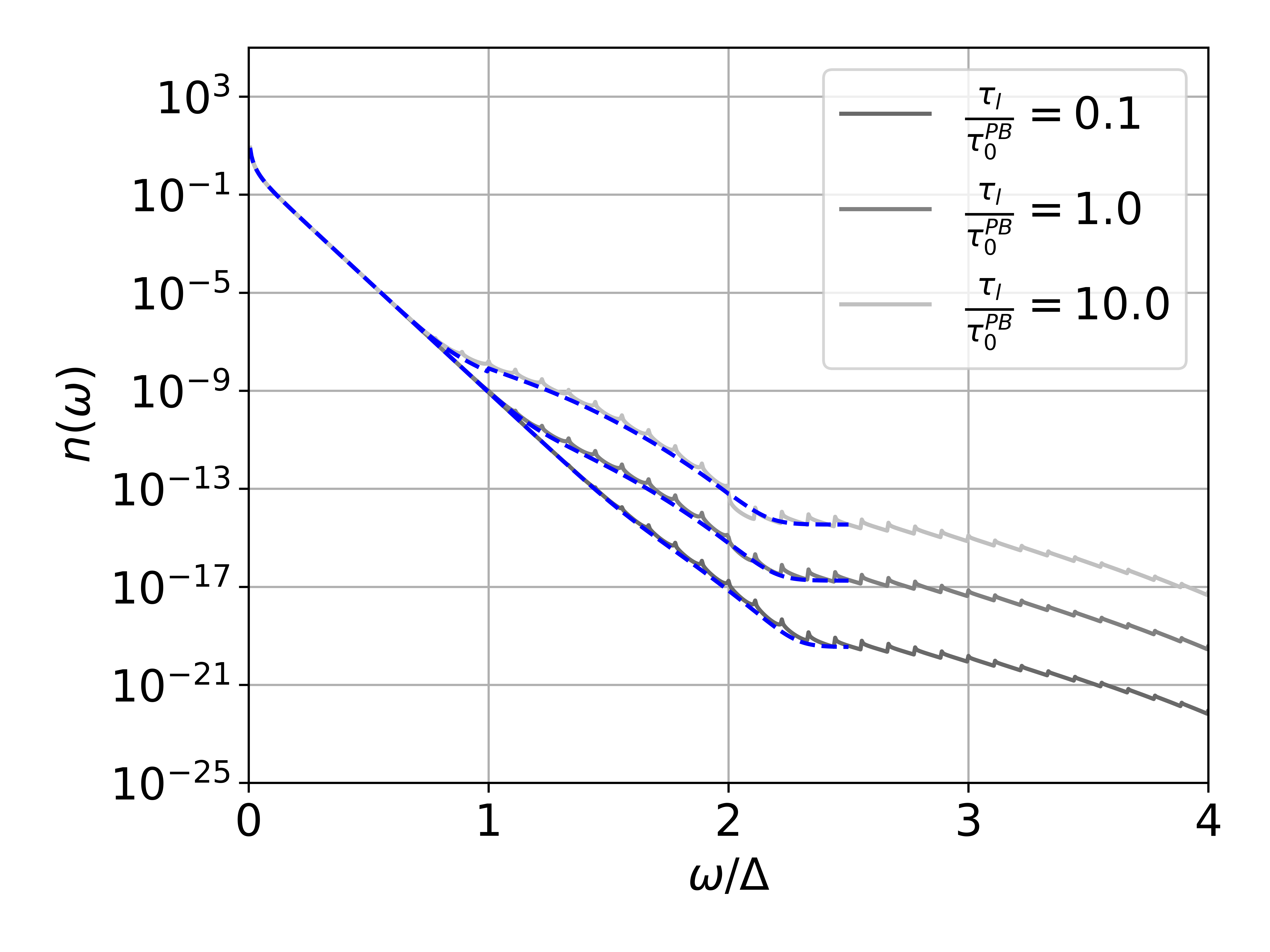}
    \caption{Phonon distribution for the system simulated in figure \ref{fig:Distributionf}. The blue dashed lines in the phonon distribution are the analytic results derived in Appendix~\ref{appendix:noneqphonons}, solid grey lines are simulation results. }
    \label{fig:Distributionn}
\end{figure}

\begin{figure}
    \includegraphics[scale=0.5]{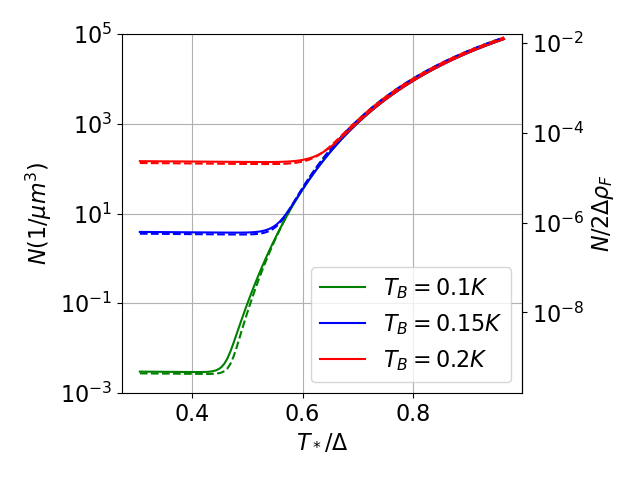}
    \includegraphics[scale=0.5]{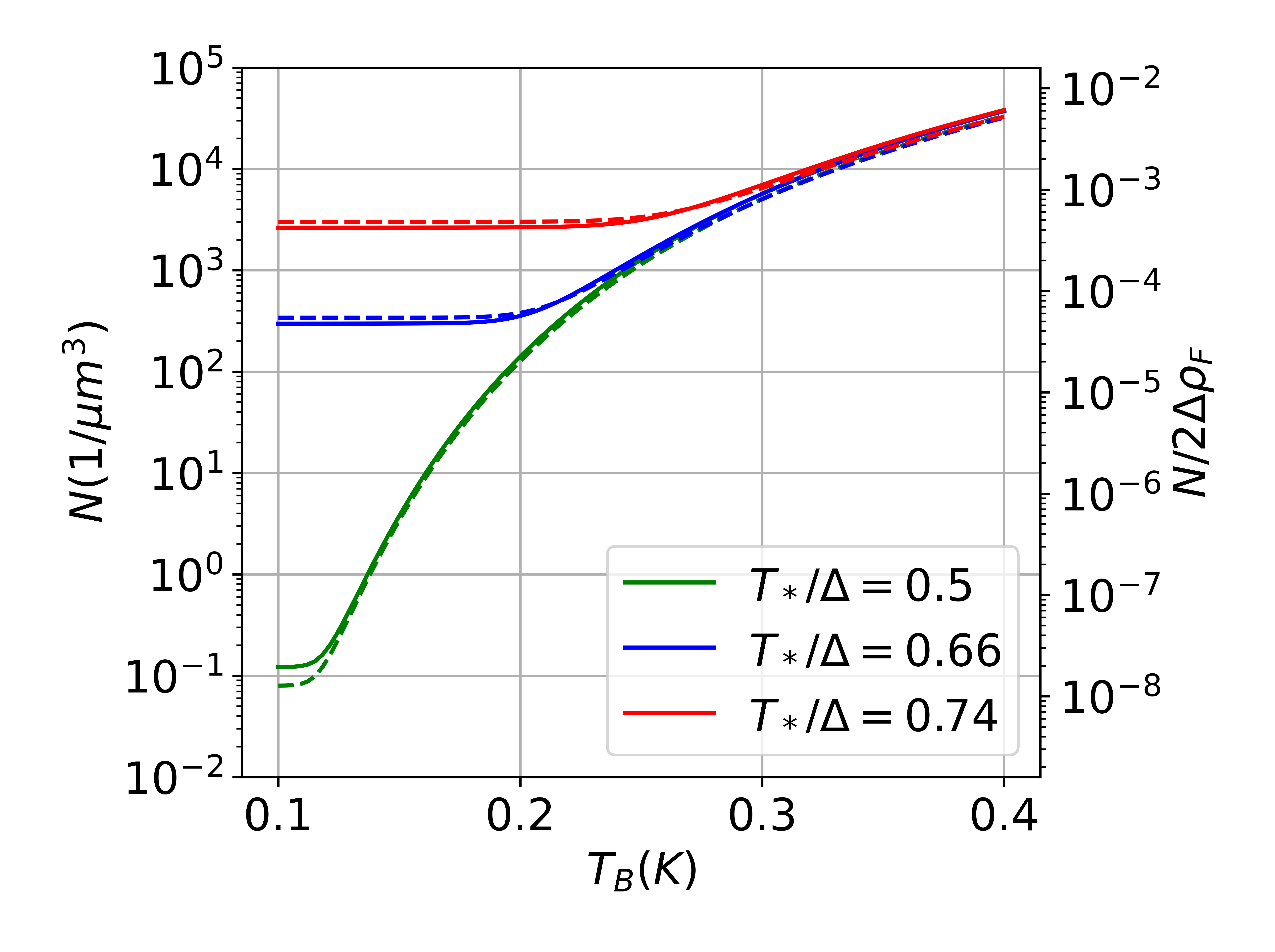}
    \caption{Quasiparticle densities calculated from the simulation (full line) and from solving Eq.~\eqref{eq:genRT} in the steady state (dashed line). The upper plot is obtained by varying the photon number for three different bath temperatures, the lower plot by varying the bath temperature for $\bar{n}=10^7,5\cdot10^7,10^8$ respectively. The thermalization time is fixed at $\tau_l=\tau_0^{PB}$, the other parameters as in Fig.~\ref{fig:Distributionf} (given in Table~\ref{tab: Param 1}). The corrections to the leading analytical results given in Appendix~\ref{appendix: Higher Order Corrections to the Quasiparticle density} have been included. }
    \label{fig: Densities}
\end{figure}

In Fig.~\ref{fig: Densities} the quasiparticle density is shown as function of $T_*$ for a few different bath temperatures $T_B$ and as function of $T_B$ for a few different phonon numbers ($T_*$). The analytical and numerical approaches give consistent results over a range of parameters relevant to experiments, with small deviations arising at low temperatures and photon numbers. These deviations are caused by the condition $\omega_0\ll T_*$ holding only weakly and we have checked that there is a closer match between the two approaches when decreasing the photon energy. In fact, for $\bar{n} \gg 1$ Eq.~\eqref{sec:Rate Equation eq:  Photon integral} is a symmetric function of $\omega_0$; therefore the leading corrections to Eq.~\eqref{sec:Shape of the quasiparticle distribution eq: Diffusion Operator} and hence to the density are of order $(\omega_0/T_*)^2$. As the magnitude of the deviations is beyond the accuracy of the considerations in the next section, we do not pursue this further.

\begin{figure}
    \centering
    \includegraphics[scale=0.5]{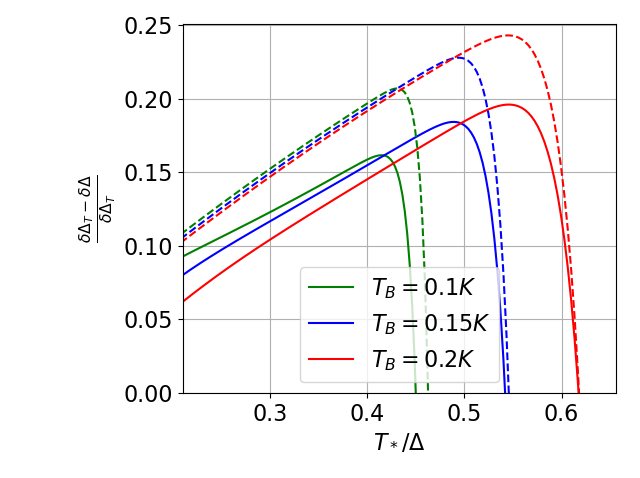}
    \caption{Suppression $\delta\Delta$ of the gap as function of $T_*$ for the same parameters used in the upper panel of Fig.~\ref{fig: Densities}. Full lines are numerical results, dashed lines correspond to the analytical formula in Eq.~\eqref{eq:gapsupp_N}. The suppression is compared to that in thermal equilibrium, $\delta\Delta_T$; a positive difference $\delta\Delta_T-\delta\Delta$ corresponds to gap enhancement. The rapid change from positive to negative coincides with the onset of large nonequilibrium quasiparticle density (c.f. Fig.~\ref{fig: Densities}).}
    \label{fig:gap}
\end{figure}

In Fig.~\ref{fig:gap} the nonequilibrium gap suppression $\delta\Delta$, Eq.~\eqref{eq:gapsupp_N}, is compared to the equilibrium one $\delta\Delta_T$ for the same set of parameters as in Fig.~\ref{fig: Densities}. Superconductivity is enhanced relative to thermal equilibrium when the difference $\delta\Delta_T - \delta\Delta$ is positive. In fact, we focus on the region of low photon number (relatively small $T_*/\Delta$) where the enhancement takes place; in this region the factor in brackets in Eq.~\eqref{eq:gapsupp_N} gives the main dependence of the gap on photon number, causing the gap to increase compared to thermal equilibrium. For larger photon numbers the quasiparticle density increases quickly (c.f. Fig.~\ref{fig: Densities}), leading to a strong suppression of the gap since  $\delta\Delta\propto N_{qp}$. The analytical estimate (dashed lines) slightly overestimates the enhancement compared to the numerics (solid lines), but both set of curves display a maximum at the temperature-dependent value of $T_*$ at which $G(T_*/\Delta)$ becomes the dominant term in Eq.~\eqref{eq:genRT}; that is, above this value quasiparticle creation from re-absorption of pair-braking phonons becomes dominant.

\section{Quality factor and resonance Frequency}
\label{sec: Quality factors}

The ac response of a superconductor depends on the quasiparticle distribution function, and the dissipative part can be strongly affected under nonequilibrium conditions, see \textit{e.g} Ref.~\cite{Nagaev.2010} and references therein. Indeed, the real part $\sigma_1$ of the ac conductivity $\sigma = \sigma_1 + i \sigma_2$ at frequency $\omega_0$ is given by
\begin{equation}\label{sec:Quality factors eq: sigma_1}
    \sigma_1=\frac{2\sigma_N}{\omega_0}\int\limits_{\Delta }^{\infty} dE\left[f(E)-f(E+\omega_0)\right]\rho (E) U^+(E,E+\omega_0)
\end{equation}
To estimate $\sigma_1$, we can use the low-energy approximation for coherence factors and density of states [cf. Eq.~\eqref{eq:Uplowen}], expand the integrand in Eq.~\eqref{sec:Quality factors eq: sigma_1} to lowest order in $\omega_0$, and use the numerical result $\int_{0}dx f'(x)/x\simeq -1.15 b_0$, where $x=(E-\Delta)/T_*$, to find
\begin{equation}
    \sigma_1 \simeq 2.3 \sigma_N \frac{\Delta}{T_*} b_0 \approx 0.77 \sigma_N \frac{N_\mathrm{qp}}{2\rho_F\Delta} \left(\frac{\Delta}{T_*}\right)^{3/2}
\end{equation}
The form on the right shows that the dependence of $\sigma_1$ on the distribution function is not only via proportionality to the quasiparticle density, due to appearance of $T_*$ in the last factor. Interestingly, as $T_*$ increases (at fixed $N_\mathrm{qp}$), dissipation decreases; this effect is due to the redistribution of quasiparticles to higher energies, where the density of states is lower, similar to the decrease in relaxation of a superconducting qubit in which a residual quasiparticle is pushed on average to higher energy in the presence of a microwave drive~\cite{Basko.2017}.

For $(\omega_0/4\Delta)^2 \ll 1$, the imaginary part $\sigma_2$ can be approximated as $\sigma_2 \simeq (\pi \sigma_N \Delta_0)/\omega_0+\delta\sigma_2$ with
\begin{align}\label{eq:ds2}
\delta\sigma_2 = & - \delta\sigma_{2,f} - \delta\sigma_{2,\Delta} \\
 \delta\sigma_{2,f} = & \frac{2\sigma_N}{\omega_0}\int\limits_{\Delta-\omega_0}^{\Delta}dE \, f(E+\omega_0)\frac{U^+(E,E+\omega_0)E}{\sqrt{\Delta^2-E^2}} \nonumber \\
\delta\sigma_{2,\Delta} = & \frac{\pi\sigma_N \delta\Delta}{\omega_0} \nonumber
\end{align}
The contribution $\delta\sigma_2$ collects all the quasiparticle effects, with $\delta\sigma_{2,f}$ accounting directly for their distribution and $\delta\sigma_{2,\Delta}$ being due to the gap suppression of Eq.~\eqref{eq:gapsupp}. Calculation of $\delta\sigma_{2,f}$ requires knowledge of the shape of distribution function within the first peak above the gap, and it is therefore beyond the description in terms of only the envelope that we have used in our analytical approach. For this reason, this term will be evaluated numerically in what follows. Still, we can give an order-of-magnitude estimate assuming $f\sim b_0$, which gives $\delta\sigma_{2,f}  \sim 2\pi\sigma_N \Delta_0 b_0/\omega_0$; this shows that the two contributions to $\delta\sigma_{2}$ can be of similar magnitude. 

Knowledge of the ac conductivity makes it possible to estimate the internal quality factor $Q_i$ and resonant frequency shift $\delta\omega_0$ of superconducting resonators. For half-wavelength, open-ended resonators made of thin superconducting film, we have~\cite{Gao.2008,Pozar.2012}
\begin{equation}\label{sec: Quality factors eq: Q_i 1}
    Q_i=\frac{\sigma_2}{\alpha\sigma_1} \simeq \frac{\sigma_N\pi\Delta_0}{\omega_0\alpha\sigma_1}\simeq \frac{\pi T_*}{2.3 \alpha \omega_0 b_0}
\end{equation}
where $\alpha$ is the kinetic inductance fraction (that is, the ratio between kinetic inductance and total inductance of the resonator), and
\begin{equation}
    \label{sec: Quality factors eq: del_omega}
  \frac{\delta \omega_0}{\omega_0}=\frac{\alpha}{2}\frac{\delta \sigma_2}{\sigma_2} \simeq -\frac{\alpha \delta\Delta}{2\Delta_0} - \frac{\alpha \omega_0 \delta\sigma_{2,f}}{2\pi\sigma_N \Delta_0}
\end{equation}
Not surprisingly, this expression resembles that for the frequency shift in superconducting qubits, in which terms originating from gap suppression and virtual transitions mediated by quasiparticle tunneling have been identified~\cite{Catelani.2011}. Note that, in contrast to the frequency shift, the first peak's shape does not affect the calculation of $Q_i$, since $\delta\sigma_2/\sigma_2 \ll 1$.

Both quality factor and frequency shift depend in general on the bath temperature $T_B$ and the photon number $\bar{n}$ through $b_0$ and $T_*$ (in particular, the quality factor scales inversely with quasiparticle density [cf. Eq.~\eqref{eq:Nqp}], but it also depends explicitly on $T_*$). For comparison to experiments, we henceforth assume that $T_B$ corresponds to the reported base temperature of the fridge. To estimate $\bar{n}$, we relate it to power absorbed by the quasiparticles $P_\mathrm{abs}$,
\begin{equation}\label{sec: Quality factors eq: nbar}
    \bar{n}=\frac{Q_i P_\mathrm{abs}}{\omega_0^2}
\end{equation}
Using this relation in Eq.~\eqref{sec: Quality factors eq: Q_i 1} we arrive at an implicit equation for $Q_i$ in terms of $P_\mathrm{abs}$. The latter is in general not a directly measurable quantity; however, for a half-wavelength resonator capacitively coupled to a transmission line we follow Ref.~\cite{Visser.2014} and relate $P_\mathrm{abs}$ to the readout power $P_\mathrm{read}$,
\begin{equation}\label{sec: Quality factors eq: P_abs}
    P_\mathrm{abs}=2 P_\mathrm{read}\frac{Q^2}{Q_iQ_c}
\end{equation}
where $Q=Q_i Q_c/(Q_i+Q_c)$ is the loaded quality factor and $Q_c$ the coupling quality factor. 
The same expression for $\bar{n}$ in terms of $P_\mathrm{read}$ is obtained using the relation between internal power and photon number, $P_\mathrm{int}= \bar{n}\omega_0^2/2\pi$, together with that between internal and readout powers, $P_\mathrm{int}=P_\mathrm{read} Q^2/\pi Q_c$~\cite{Gao.2008}. Equations \eqref{sec: Quality factors eq: Q_i 1}, \eqref{sec: Quality factors eq: nbar}, and \eqref{sec: Quality factors eq: P_abs} make possible a self-consistent calculation of photon number and internal quality factor as a function of readout power. In the following we solve these equations explicitly in two regimes.

The approach just described simplifies considerably in the limit $Q_i \gg Q_c$, in which Eqs.~\eqref{sec: Quality factors eq: nbar} and \eqref{sec: Quality factors eq: P_abs} reduce to $\bar{n}= 2 Q_c P_\mathrm{read}/\omega_0^2$, independent of $Q_i$. Then $T_*$ can be calculated from its definition, Eq.~\eqref{eq:Tstardef}, and $b_0$ by solving Eq.~\eqref{eq:Normalization eq}. In what follows, we denote with $T_{*,0}$ the value of $T_*$ obtained under the assumption $Q_i \gg Q_c$, namely
\begin{equation}\label{eq: T_star_0}
\frac{T_{*,0}}{\Delta} \equiv  \left[\frac{105}{64}\left(\frac{T_c}{\Delta}\right)^3c_{Phot}^{QP}\tau_0 \frac{2P_\mathrm{read}Q_c}{\Delta^2}\right]^{1/6}    
\end{equation} 
If we further consider the low phonon temperature/high photon number regime, $T_B \ll T_B^*$, we find that the quality factor is given by
\begin{equation}\label{eq:Qi0}
Q_{i,0}  = \frac{\gamma_0 \Delta}{\alpha \omega_0} \frac{\tau_0^{PB}}{\tau_l}\left(\frac{\Delta}{T_{*,0}}\right)^3 e^{\sqrt{14/5}(\Delta/T_{*,0})^3}
\end{equation}
with $\gamma_0 = \pi 2^{1/3}5\sqrt{7}(2.1)^2/(2.3\times 3^{3/2})\simeq 19.3$. Therefore in this case the quality factor is a decreasing function of readout power and does not depend on $T_B$. The fast decrease of quality factor with increasing power is due to the large increase in the number of quasiparticles with photon number, as discussed in Sec.~\ref{sec:genRT}.

The high phonon temperature/low photon number regime $T_B \gtrsim T_B^*$ also offers significant simplification, since in that case we can relate $b_0$ to the (thermal) quasiparticle density using Eq.~\eqref{eq:Nqp}, and rewrite Eq.~\eqref{sec: Quality factors eq: Q_i 1} as
\begin{equation}
    Q_i = \frac{2.1\sqrt{2} \pi}{2.3 \alpha} \frac{\Delta}{\omega_0}\left(\frac{T_*}{\Delta}\right)^{3/2} \left(\frac{N_\mathrm{qp}^{T_B}}{2\rho_F \Delta}\right)^{-1}
\end{equation}
where the superscript $T_B$ in $N_\mathrm{qp}^{T_B}=2\rho_F\Delta \sqrt{2\pi T_B/\Delta} \times e^{-\Delta/T_B}$ denotes the thermal equilibrium density at temperature $T_B$, which is independent of $\bar{n}$.
This formula, using Eqs.~\eqref{eq:Tstardef}, \eqref{sec: Quality factors eq: nbar} and \eqref{sec: Quality factors eq: P_abs}, leads to a quadratic equation for $Q_i$ whose solution can be written in the form
\begin{align}\label{eq:Qi_highT}
& Q_i = \\ & \sqrt{\left(\frac{Q_c}{2}\right)^2 + \left(\frac{2.1\sqrt{2}\pi \Delta}{2.3 \alpha \omega_0}\right)^2 \left(\frac{T_{*,0}}{\Delta}\right)^3 \left(\frac{N_\mathrm{qp}^{T_B}}{2\rho_F\Delta}\right)^{-2}}-\frac{Q_c}{2}
\nonumber
\end{align}
In this regime, $Q_i$ is an increasing function of readout power (through $T_{*,0}$) and depends exponentially on $T_B$ (through $N_\mathrm{qp}^{T_B}$). As remarked above, the increase of quality factor with readout power can be traced to the redistribution of the quasiparticles to higher energies, where the density of states is lower. Note that if $Q_i \gg Q_c$ holds up to temperatures of order $T_B^*$ or higher, Eqs.~\eqref{eq:Qi0} and \eqref{eq:Qi_highT} together capture the temperature and power dependence of $Q_i$ at all temperatures; we do not investigate here the case in which this condition is not satisfied.

So far we have assumed that the internal quality factor is determined by the energy absorbed by quasiparticles. More generally, extrinsic (that is, non-quasiparticle) mechanisms such as dielectric losses can contribute to the total internal quality factor $Q_{i,\mathrm{tot}}$; collecting those contributions into $Q_{i,\mathrm{ext}}$ we have
\begin{equation} \label{eq:Qtot}
1/Q_{i,\mathrm{tot}} = 1/Q_i+1/Q_{i,\mathrm{ext}}
\end{equation}
where as before $Q_i$ denotes the quasiparticle part. For $T_B \gtrsim T_B^*$, $Q_i$ is given by Eq.~\eqref{eq:Qi_highT} with the replacement $1/Q_c \to 1/Q_c + 1/Q_{i,\mathrm{ext}}$ and is therefore unchanged (at leading order) if $Q_{i,\mathrm{ext}} \gg Q_c$.

\subsection{Comparison to experiments}
\label{sec:expcomp}

\begin{figure*}
    \centering
    \includegraphics[scale=.5]{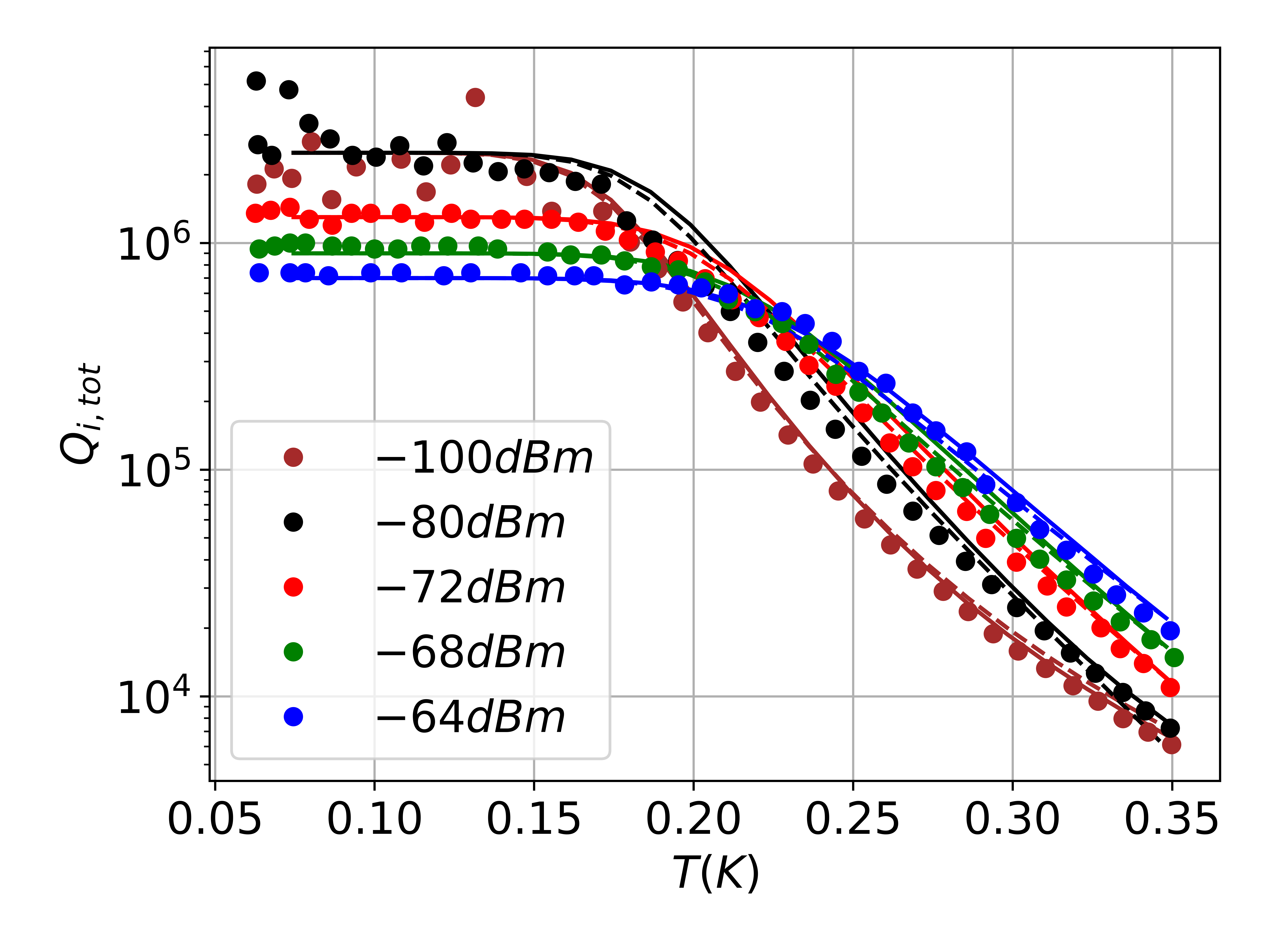}
    \includegraphics[scale=.5]{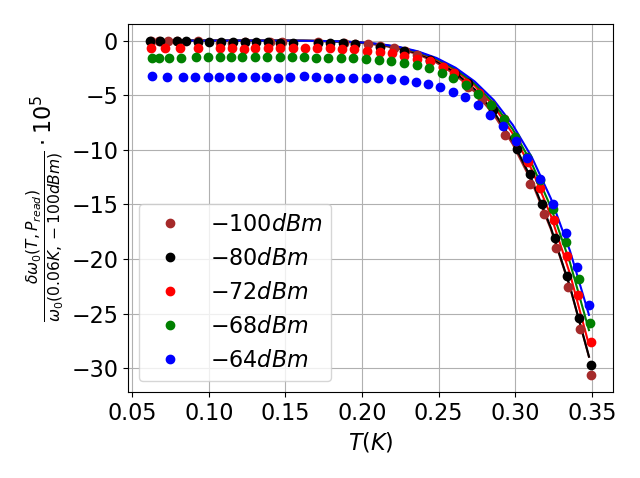}
    \caption{Internal quality factor $Q_{i,\mathrm{tot}}$ vs temperature for different readout powers (left) and deviation of the resonance frequency from its experimentally determined low temperature and low readout power value $\delta\omega_0(T,P_\mathrm{read})=\omega_0(T,P_\mathrm{read})-\omega_0(T=0.06\,\mathrm{K},P_\mathrm{read}=-100\,\mathrm{dBm})$ in units of $10^{-5}\omega_0$ for different readout powers (right). Experimental data from \cite{Visser.2014} are displayed with circles (we omit the data for $P_\mathrm{read} = -90\,$dBm for clarity of presentation), results from the numerics described in Sec.~\ref{sec:Rate Equation eq: Numerical approach} with full lines, and results from the analytical formulas [Eqs.~\eqref{eq:Qi0}-\eqref{eq:Qtot}] with dashed lines. The analytical curve for the lowest readout power assume thermal equilibrium (see Appendix~\ref{appendix: Simulation Parameter}).}
    \label{fig:Quality factors}
\end{figure*}

We now proceed to compare our theoretical findings to the measurements of the temperature dependencies of the quality factor and resonant frequency for different readout powers reported in Ref.~\cite{Visser.2014}.
In that work, a temperature-independent plateau in the quality factor at low temperatures is observed, which qualitatively agrees with the result in Eq.~\eqref{eq:Qi0}. In fact, by comparing experiments to numerical calculations the authors of Ref.~\cite{Visser.2014} suggest the plateau to be explainable by the nonequilibrium steady-state solution to the kinetic equations.
However, estimating the value of the plateau $Q_{i,0}$, Eq.~\eqref{eq:Qi0}, using the parameters in Table~\ref{tab: Param Ex}, we find much larger values than measured experimentally, see Table~\ref{tab:Limiting Quality factors}. The discrepancy cannot be due to approximations being used: while in particular the assumption $\omega_0\ll T_*$ holds only weakly, for the highest readout power comparison with numerical results (see Fig.~\ref{fig: Densities}) shows that our analytical expression underestimates the quasiparticle density by a factor smaller than 2; then the quality factor could be overestimated by the same factor, but the estimated quality factor is three orders of magnitude larger than the measured one~\footnote{We stress here that in discretizing the kinetic equations attention must be paid as to avoid introducing an unphysical quasiparticle source (cf.~\cite{Note2}); in numerical calculations this would cause saturation of the quality factor to levels lower than our estimates provide.}. This indicates that the plateau is not due to the quasiparticles being driven out of equilibrium by the resonator's photons, but either to some other driving mechanism and/or to extrinsic relaxation channels. The measured power dependence of the quality factor is qualitatively opposite to that typically expected in the presence of two-level systems~\cite{Macha}, so they are unlikely to be the cause of the plateau. Moreover, the low temperature saturation of the quasiparticle lifetime reported in Ref.~\cite{Visser.2014} (for a different sample and for a narrower range of power) point to the presence of a second driving mechanism, a situation that deserves further study. Nonetheless, assuming for simplicity an extrinsic mechanism, to fit the experimental data we use Eq.~\eqref{eq:Qtot} and find the values of $Q_{i,\mathrm{ext}}$ given in Table~\ref{tab:Limiting Quality factors}. Results from the analytic formula and the numerics are compared to experimental data for quality factor and resonance frequency in Fig.~\ref{fig:Quality factors}; in the numerical calculations, the experimentally measured (total) internal quality factor has been used in Eqs.~\eqref{sec: Quality factors eq: nbar}-\eqref{sec: Quality factors eq: P_abs} to obtain the photon number. 
At temperatures $T\gtrsim 0.25\,$K we find good agreement between theory and experiment. The fitted value of $\tau_0$ is shorter than an estimate derived from neutron scattering data ~\cite{Kaplan.1976} but consistent with other experimental estimates, see \textit{e.g.} Ref.~\cite{Chi.1979}. This further validates our approach, especially since from the analytical expressions it is evident that $\tau_0$ enters into the quality factor via the product $c^{QP}_{Phot} \tau_0 \bar{n}$, and therefore inaccuracies in the estimates of $c^{QP}_{Phot}$ and/or $\bar{n}$ (equivalently, $P_\mathrm{read}$) could affect the extracted value of $\tau_0$.

The disagreement between theory and experiment for the frequency shift at low temperature and readout power larger than -80\,dBm could perhaps be due to the same driving and/or extrinsic mechanisms responsible for the saturation of the low-temperature quality factor. In fact, it is suggested in Ref.~\cite{Semenov} that the depairing effect of a microwave drive modifies the density of states in such a way to cause a negative frequency shift proportional to the power; however, the corresponding influence on the quality factor was not analyzed.
Finally, we note that in agreement with the discussion after Eq.~\eqref{eq:ds2}, based on our numerics about 40 to 45\,\% of the calculated frequency shift originates from the gap suppression, comparable to the direct contribution due to the nonequilibrium distribution.

\begin{table}[!bt]
\begin{tabular}{|c|c|c|c|c|c|c|}
\hline
$c^{QP}_{Phot}$ & $\tau_0$ & $\Delta_0$ &$\omega_0$ & $\tau_{0}^{PB}$ & $\tau_l$ & $\alpha$ \\
\hline
$0.06\,$Hz & $63\,$ns & $189\,\mu$eV &  22$\Delta_T$/189 & $40\,$ps & 170\,ps & 0.13 \\
\hline
\end{tabular}
\caption{Parameters used for comparison between theory in this work and experiments of Ref.~\cite{Visser.2014}. 
The estimation of $c^{QP}_{Phot}$, $\Delta_0$, and $\alpha$ is discussed in Appendix~\ref{appendix: Simulation Parameter}. For the phonon thermalization time $\tau_l$ and the coupling quality factor $Q_c = 20100$ we use the same values as in Ref.~\cite{Visser.2014}. In the numerical calculations, the gap is fixed at its thermal equilibrium value $\Delta_T$, the discretization step size is $h=\Delta_T/189$, and the energy is truncated at $E_\mathrm{max}=10\Delta_T$. The time $\tau_0$ is used as a free fit parameter, while $\tau_0^{PB}$ follows from it via Eq.~\eqref{sec:Rate Equation eq: Relation Lifetimes}.}
\label{tab: Param Ex}
\end{table}

\begin{table}
    \centering
    \begin{tabular}{|c|c|c|c|c|c|c|}
    \hline
$P_\mathrm{read}$ (dBm) & -100 &-90
&-80&-72&-68&-64 \\
\hline
$T_{*,0}/\Delta$ & $0.12$ &$0.18$
&$0.26$&$0.36$&$0.42$&$0.49$ \\
\hline
$Q_{i,0}/10^6$& &$10^{124}$
&$10^{38}$&$10^{13}$&$10^{7}$&$10^3$ \\
\hline
$Q_{i,\mathrm{ext}}/10^6$&2.5
&2.5
&2.5&1.3&0.9&0.7 \\
\hline
    \end{tabular}
    \caption{For each readout power used in Ref.~\cite{Visser.2014}, we calculate $T_{*,0}$ using Eq.~\eqref{eq: T_star_0} and the (theoretical) low-temperature quality factors $Q_{i,0}$ using Eq.~\eqref{eq:Qi0} (except for the lowest power, since in that case $T_{*,0}\approx \omega_0$, while the theory is valid for $T_{*,0} >  \omega_0$). The experimental low-temperature quality factors $Q_{i,\mathrm{ext}}$ are assumed to be due to extrinsic (non-quasiparticle) mechanisms (see text).}
    \label{tab:Limiting Quality factors}
\end{table}

\section{Summary}
\label{sec:conclusions}

In this work we investigate the quasiparticle distribution function in superconducting resonators in the regime of low quasiparticle density in the presence of a large number of low energy ($\omega_0\ll\Delta$) photons and of a low temperature ($T_B \ll \Delta$) phonon bath.
In the steady state, we present approximate analytical solutions to the kinetic equations governing the dynamics of the quasiparticle and phonon distribution functions. The shape of the quasiparticle distribution function -- that is, its functional dependence on energy -- is determined by interplay between absorption and emission of photons and phonons, see Sec.~\ref{sec: Shape of the quasiparticle distribution}, and has a typical width $T_*$ above the gap [see Eq.~\eqref{eq:Tstardef} and Ref.~\cite{Basko.2017}]. The overall normalization and hence the quasiparticle density are controlled by the balance between recombination by phonon emission and generation, see Sec.~\ref{sec: Quasiparticle Density}; due to the photons driving the quasiparticles out of equilibrium, the generation is due not only to thermal phonons but also to nonequilibrium phonons emitted by quasiparticles of sufficiently high energy. The density dynamics then follows from the generalized Rothwarf-Taylor model of Sec.~\ref{sec:genRT}; for $T_B > T_B^* = T_*^3/\Delta^2$ the steady-state density is approximately the same as in thermal equilibrium, while it is much larger than that at lower bath temperatures. The analytical results are validated by comparison with numerical calculations.

Our results enable us to calculate the dependence on temperature and readout power of the internal quality factor and (numerically) of the frequency shift in thin-film resonators, see Sec.~\ref{sec: Quality factors}. In contrast to previous suggestions~\cite{Goldie.2012,Visser.2014}, we find that the quasiparticle distribution driven out of equilibrium by the resonator's photons cannot explain the experimental data of Ref.~\cite{Visser.2014} at low temperature, while it quantitatively describes them above about 0.25~K, see Fig.~\ref{fig:Quality factors}.

\acknowledgments

We gratefully acknowledge D. Basko for interesting discussions. This work was supported in part by the German Federal Ministry of Education and Research (BMBF), funding program ``Quantum technologies -- from basic research to market'', project QSolid
(Grant No. 13N16149).

\appendix

\section{Quasiparticle and phonon lifetimes}
\label{appendix:Lifetimes}

In Sec.~\ref{sec:Lifetimes} a number of lifetimes have been introduced to qualitatively discuss the behaviour of the quasiparticle distribution function out of equilibrium. In this appendix, we give their precise definitions in terms of the non-equilibrium quasiparticle and phonon distributions, as obtained by inspecting the structure of the kinetic equations. As in Sec.~\ref{sec:Lifetimes}, we consider first the phonon lifetimes.

The lifetime of a phonon of energy $\omega$ against pair breaking is [cf. the last term in curly bracket in the second integral in Eq.~\eqref{sec:Rate Equation eq: Phonon Rate Equation}]
\begin{equation}
\begin{split}
    \frac{1}{\tau^{Phon}_{PB}(\omega)}=\frac{1 }{\pi \Delta_0 \tau^{PB}_0}& \int\limits_{\Delta}^{\omega-\Delta}\!dE\,\rho(E) U^+(E,\omega-E) \\
    &\times\left[1- f(\omega-E)\right]\left[1- f(E)\right].
\end{split}
\end{equation}
Approximating the Pauli-blocking factors as $[1-f(E)] \simeq 1$, as done throughout this paper, the dependence of $1/\tau_{PB}^{Phon}$ on $\omega$ can be expressed in terms of the spectral density $S_+$ of Refs.~\cite{Houzet.2019,Glazman.2021} as
\begin{equation}\label{eq:tPBPhonS}
    \frac{1}{\tau^{Phon}_{PB}(\omega)}=\frac{\Delta}{\pi \Delta_0 \tau^{PB}_0} S_+ (\omega/\Delta) \, ,
\end{equation}
where
\begin{equation}\label{eq: Spectral Density S+}
    S_+(x) = (x+2) E\left(\frac{x-2}{x+2}\right) - \frac{4x}{x+2}K\left(\frac{x-2}{x+2}\right)
\end{equation}
with $E$ and $K$ the complete elliptic integrals of the second and first kind, respectively. Note that $S_+(x) = 0$ for $x<2$, $S_+(x) \approx x$ for $x \gg 2$, and $S_+(x) \approx \pi \left[1+ (x-2)/4\right]$ for $x-2 \ll 2$. The last expression shows that the pair-breaking lifetime approaches $\tau_0^{PB}$ at low temperature and only weakly depends on energy for $\omega \gtrsim 2\Delta$.

The lifetime of a phonon against being absorbed by a quasiparticle is [cf. the last term in curly bracket in the first integral in Eq.~\eqref{sec:Rate Equation eq: Phonon Rate Equation}]
\begin{align}
    \frac{1}{\tau^{Phon}_{abs}(\omega)}=\frac{2 }{\pi \Delta_0 \tau^{PB}_0}\int\limits_{\Delta}^{\infty}&\!dE\, \rho(E) U^-(E,E+\omega)\nonumber\\
    &\times f(E)\left[1-f(E+\omega)\right].
\end{align}
For $f \ll 1$, one can show that the right-hand side is bounded by $1/\pi \tau_0^{PB}$ times the normalized quasiparticle density $N_\mathrm{qp}/2\rho_F \Delta$. Since at low temperature the latter is much smaller than 1, we have $\tau_{abs}^{Phon} \gg \tau_0^{PB}$.

The last phonon lifetime we take into account is the thermalization time $\tau_l$. For of a phonon of energy $2\Delta$ whose mean free path against pair breaking $s\tau_0^{PB}$ (with $s$ the speed of sound) is of a similar order as the film thickness $d$, $\tau_l$  can be estimated to be~\cite{Kaplan.1979,Eisenmenger.1976,Chang.1978}
\begin{equation}\label{Thermalization Time}
\tau_l \approx \frac{4 d}{\eta s}
\end{equation}
where $\eta$ is the direction-averaged transmission coefficient between film and substrate. For aluminium on sapphire, the speed of sound  and transmission coefficient are approximately $s\simeq 3.3$-$6.7\,$km/s and $\eta\simeq 0.2$~\cite{Kaplan.1979}.
For the resonator of Ref.~\cite{Visser.2014} considered in Sec.~\ref{sec: Quality factors}, this expression gives $\tau_l\approx180$-$360\,$ps.
In Ref.~\cite{Visser.2014} the value $\tau_l=170\,$ps was used, and to simplify comparison to their results we also use this latter value in our analysis. Our results do not depend on the exact value of the thermalization time at energies $\omega<2\Delta$, if the condition $\tau^{Phon}_{abs}\gg \tau_l$ holds, as is the case for both choices of $\tau_l$.
We note that some authors~\cite{Kozorezov.2004,Kozorezov.2004b} suggest the thermalization time of phonons with $\omega<2\Delta$ to be limited by total internal reflection and, for sufficiently smooth interfaces, to be several orders of magnitude longer than the one at energies above $2\Delta$.
We assume this not to be the case and use the thermalization time at energy $\omega=2\Delta$ for all energies.

We now turn to the quasiparticle lifetimes. The scattering lifetime $\tau^{qp}_{s,t} $ of a quasiparticle due to interaction with photons has three contributions
\begin{equation}
    \frac{1}{\tau^{qp}_{s,t}} = \frac{1}{\tau^{qp}_{sp,t}} +\frac{1}{\tau^{qp}_{st,t}} + \frac{1}{\tau^{qp}_{abs,t}}
\end{equation}
originating respectively from spontaneous emission, stimulated emission, and absorption of a photon. They are given respectively by [cf. the second terms in curly brackets in Eq.~\eqref{sec:Rate Equation eq:  Photon integral}]
\begin{equation}
    \frac{1}{\tau^{qp}_{sp,t}}=c^{QP}_{Phot}U^+(E,E-\omega_0)\left[1-f(E-\omega_0)\right]
\end{equation}
\begin{equation}
    \frac{1}{\tau^{qp}_{st,t}}=c^{QP}_{Phot}\bar{n}U^+(E,E-\omega_0)\left[1-f(E-\omega_0)\right]
\end{equation}
\begin{equation}
    \frac{1}{\tau^{qp}_{abs,t}}=c^{QP}_{Phot}\bar{n}U^+(E,E+\omega_0)\left[1-f(E+\omega_0)\right]
\end{equation}

The scattering lifetime $\tau^{qp}_{s,n}$ due to interaction with phonons has also three contributions
\begin{equation}
    \frac{1}{\tau^{qp}_{s,n}} = \frac{1}{\tau^{qp}_{sp,n}} +\frac{1}{\tau^{qp}_{st,n}} + \frac{1}{\tau^{qp}_{abs,n}}
\end{equation}
accounting for spontaneous phonon emission [cf. the last term in curly brackets in Eq.~\eqref{sec:Rate Equation eq: Gamma^Phon_sp}]
\begin{equation}
   \frac{1}{\tau^{qp}_{sp,n}}=\frac{1}{\tau_0 T_c^3}\int\limits_{0}^{E-\Delta}\!d\omega\,\omega^2 U^-(E,E-\omega)\left[1-f(E-\omega) \right]
\end{equation}
stimulated phonon emission [cf. the last term in curly brackets in the second integral of Eq.~\eqref{sec:Rate Equation eq: Stimulated Phonon integral}]
\begin{equation}
    \frac{1}{\tau^{qp}_{st,n}}=\frac{1}{\tau_0 T_c^3}\int\limits_{0}^{E-\Delta}\!d\omega\, \omega^2 U^-(E,E-\omega)n(\omega)\left[1-f(E-\omega) \right]
\end{equation}
and phonon absorption [cf. the last term in curly brackets in the first integral of Eq.~\eqref{sec:Rate Equation eq: Stimulated Phonon integral}]
\begin{equation}
    \frac{1}{\tau^{qp}_{abs,n}}=\frac{1}{\tau_0 T_c^3}\int\limits_{0}^{\infty}d\omega\, \omega^2 U^-(E,E+\omega)n(\omega)\left[1-f(E+\omega) \right]
\end{equation}
In addition, the lifetime of a quasiparticle against recombination is [cf. Eq.~\eqref{sec:Rate Equation eq: Gamma^Phon_re}]
\begin{equation}
    \frac{1}{\tau^{qp}_{r}}=\frac{1}{\tau_0 T_c^3}\int\limits_{E+\Delta}^{\infty}d\omega\,\omega^2 U^+ (E,\omega-E) f(\omega-E)\left[1+n(\omega)\right]
\end{equation}
For a system in equilibrium, these lifetimes (except those due to photons) are discussed in Ref.~\cite{Kaplan.1976}. When $f\ll 1$, the only lifetimes with significant dependence on the quasiparticle distribution function are $\tau^{Phon}_{abs}$ and $\tau^{qp}_r$: since the rates are proportional to $f$, the times are inversely proportional to the quasiparticle density. In fact, in Appendix~\ref{appendix: Higher Order Corrections to the Quasiparticle density} we study the dependence of the recombination coefficient $R$ [that is, the recombination rate for the quasiparticle density, see Eq.~\eqref{eq:genRT}] on $T_*$, while as discussed in Sec.~\ref{sec:Shape of the quasiparticle distribution Phonon trapping} the effect of $\tau^{Phon}_{abs}$ on the quasiparticle distribution can be neglected.

\section{Finite phonon temperature}
\label{appendix: Laplace approximation}

In this appendix we present in some detail the derivation of the results discussed in Sec.~\ref{sec:finteTphonons}. As mentioned there, we want to find the energy $E_*$ below which neglecting the effect of thermal phonons is justified, in which case the formulas of Sec.~\ref{sec:zeroTphonons} can be used, and to find $E_*$ we need an estimate for $St_{st}^{Phon}$, Eq.~\eqref{sec:Rate Equation eq: Stimulated Phonon integral}, at $E=E_*$. In the integrands in that equation, we assume as usual $f \ll 1$ at all energies; then as long as factors other than $n(\omega)=n_T(\omega,T_B)$ grow with $\omega$ slower than exponentially, the integral is determined by the integration in the interval around $\omega \sim T_B$. This is the case for all terms except that proportional to $f(E_*-\omega)$, which increases faster than exponentially; for this term, the integral can be estimated as follow: we consider a frequency $\omega_*$ such that $\omega_* \gg T_B$ and $\omega_M-\omega_* \gg \tilde{T}_*$, with $\omega_M \gg T_B$ to be defined below (here we assume $E_* \gtrsim 2\Delta$; for $E_* \lesssim 2\Delta$ the condition reads $\omega_M-\omega_* \gg T_*$, with $\omega_M$ having different definitions in the two cases). The contribution to the integral from the interval $0<\omega < \omega_*$ can be estimated together with the other terms in Eq.~\eqref{sec:Rate Equation eq: Stimulated Phonon integral}, while that from $\omega_*<\omega< E_* - \Delta$ is to be considered separately.

Let us start with the low-frequency contribution, $\omega < \omega_*$, which we denote with $St_{st,low}^{Phon}$. Then since $n$ restricts $\omega$ to be of order $T_B \ll E_*$, while the (envelope of the) quasiparticle distribution function $f$ varies at most over a scale $T_* \gg T_B$, we can approximate $U^-(E_*,E_* \pm \omega) \approx 1$, perform a series expansion for the distribution functions, and push the integration limit from $\omega_*$ to infinity to find
\begin{equation}
   St_{st,low}^{Phon} \approx \frac{1}{\tau_0 T_c^3}\int_0^\infty d\omega \, \omega^4 \frac{\partial^2 f}{\partial E^2} \bigg|_{E=E_*} n(\omega,T_B) \, .
\end{equation}
According to Eq.~\eqref{sec: Shape of the quasiparticle distribution eq: High Energy Solution approximate}, for $(E_*-\Delta)/\tilde{T}_* \gg 1$ we have 
\begin{equation}
    \frac{\partial^2 f}{\partial E^2} \bigg|_{E=E_*} \approx \frac{1}{\tilde{T}_*^2}\left(\frac{E_*-\Delta}{\tilde{T}_*}\right)^3 f(E_*)
\end{equation}
and therefore
\begin{equation}
   St_{st,low}^{Phon} \approx \frac{24\zeta(5)}{\tau_0 T_c^3} \left(\frac{T_B}{\tilde{T}^*}\right)^5 \left(E_*-\Delta\right)^3 f(E_*) \, ,
\end{equation}
with $\zeta$ denoting the Riemann zeta function. Comparing this expression to Eq.~\eqref{eq:StPhotHighE}, it is clear that the low-frequency contribution can be neglected for $T_B \ll \tilde{T}_*$. The similar calculation for the case $E_* \lesssim 2\Delta$ is slightly more complex, as one needs to use the approximation in Eq.~\eqref{sec:Shape of the quasiparticle distribution eq: Low Energy approximation U_-} for $U^-$; the resulting condition reads $(E_*-\Delta) T_B^5/T_*^6 \ll 1$.

We now turn to the high-frequency contribution
\begin{equation}\label{eq:St_sthi}
  St_{st,hi}^{Phon} \approx \frac{1}{\tau_0T_c^3}\int_{\omega_*}^{E_*-\Delta} d\omega \, \omega^2 f(E_*-\omega) e^{-\omega/T_B} \, ,
\end{equation}
where the approximations employed are $U^- \approx 1$ and $n(\omega,T_B) \approx e^{-\omega/T_B}$. For $E_* - \omega > 2\Delta$, we can use Eq.~\eqref{sec: Shape of the quasiparticle distribution eq: High Energy Solution approximate} for $f$ and estimate the integral using Laplace's method; indeed, the argument of the exponential has a maximum at the frequency $\omega_M = E_*-\Delta - \tilde{T}_* (\tilde{T}_*/T_B)^{2/3}$, and $E_*-\omega_M > 2\Delta$ if $T_B/\Delta < (T_*/\Delta)^3$. Similarly, for $E_*-\omega < 2\Delta$ one can use the asymptotic approximation for the result in Eq.~\eqref{sec: Shape of the quasiparticle distribution eq: Low Energy Solution high energy} to find $\omega_M = E_*-\Delta-T_* (T_*/T_B)^{1/2}$, and $E_*-\omega_M < 2\Delta$ if $(T_*/\Delta)^3 < T_B/\Delta \ll T_*/\Delta$ [we note that the algebraic prefactor is different in this case, as $U^-$ takes approximately the form given in Eq.~\eqref{sec:Shape of the quasiparticle distribution eq: Low Energy approximation U_-}].
The estimate of the integral in Eq.~\eqref{eq:St_sthi} is then obtained by evaluating the prefactor at $\omega= \omega_M$, expanding the argument of the exponential up to second order around $\omega_M$, and finally performing the resulting Gaussian integral; here the integration limits can be extended to infinity, since the width of the Gaussian peak is of order $\tilde{T}_*(T_B/\tilde{T}_*)^{1/6}$ for $E_*>2\Delta$ and $T_* (T_B/T_*)^{1/4}$ for $E_* < 2\Delta$. 

We can now use the estimate thus found for $St^{Phon}_{st}$ in Eq.~\eqref{sec: Shape of the quasiparticle distribution eq: Definition E_*} together with Eq.~\eqref{eq:StPhotHighE} to arrive at the following equation for $E_*>2\Delta$:
\begin{equation}\label{sec:Shape of the quasiparticle distribution eq: Implicit Equation tilde(x)_*}
    4\sqrt{3\pi}(\tilde{x}_*-\tilde{\chi}^{2/3})^2e^{\frac{3}{5}\tilde{\chi}^{5/3}+\frac{2}{5}\tilde{x}_*^{5/2}-\tilde{\chi}\tilde{x}_* }=\tilde{x}_*^{9/4} \tilde{\chi}^{2/3}
\end{equation}
where $\tilde{x}_*=(E_*-\Delta)/\tilde{T}_*$ and $\tilde{\chi}=\tilde{T}_*/T_B$. For $E_* < 2\Delta$, the term $St^{Phot}$ in Eq.~\eqref{sec: Shape of the quasiparticle distribution eq: Definition E_*} can be expressed as the opposite of the first term on the right-hand side of Eq.~\eqref{sec:Shape of the quasiparticle distribution eq: Low Energy approximation Phonon term}, and we similarly find the equation
\begin{equation}\label{sec:Shape of the quasiparticle distribution eq: Implicit Equation x_*}
105\sqrt{\pi}(x_*+\sqrt{\chi})(x_*-\sqrt{\chi})^2e^{\frac{2}{3}\chi^{3/2}+\frac{1}{3}x_*^3-x_*\chi }=64 x_*^{3} \chi^{3/4}
\end{equation}
with $x_*=(E_*-\Delta)/T_*$ and $\chi=T_*/T_B$. These equations can be solved approximately by expanding up to second order, both in the prefactors and in the argument of the exponential, around $\tilde{x}_* = \tilde{\chi}^{2/3}$ ($x_* = \sqrt{\chi}$), to find for $T_B \ll T_B^*$
\begin{align}
\label{sec: App eq: Approximate E_* High E}
&E_* \simeq \Delta + \tilde{T}_*(\tilde{T}_*/T_B)^{2/3}\\
&+\tilde{T}_*\sqrt{\frac{4}{3}\left(\frac{T_B}{\tilde{T}_*} \right)^{1/3}W\left[\frac{\sqrt{3}}{16\sqrt{\pi}}\left(\frac{\tilde{T}_*}{T_B}\right)^{5/2} \right]} \gtrsim 2\Delta \nonumber
\end{align}
where $W$ is the Lambert or product logarithm function with the asymptotic behavior $W(x) \approx \ln x - \ln \ln x$ for $x\gg 1$, while for $T_B^* \lesssim T_B \ll T_*$ we get  
\begin{align}
\label{sec: App eq: Approximate E_* Low E}
     &E_* \simeq \Delta + T_* (T_*/T_B)^{1/2} \\
     &+T_*\sqrt{\left(\frac{T_B}{T_*}\right)^{1/2}W\left[\frac{32}{105\sqrt{\pi}}\left(\frac{T_*}{T_B}\right)^{9/4} \right]}\lesssim 2\Delta\, .\nonumber
\end{align}
In Sec.~\ref{sec:finteTphonons} we have reported for simplicity only the leading terms for $E_*$.

By construction, the energy $E_*$ denotes that energy at which the effects of absorption and stimulated emission of phonons become comparable to that of photons, so that neglecting $St_{st}^{Phon}$ is justified only below $E_*$. Conversely, we now show that for $E>E_*$ we can neglect the photons, and the quasiparticle distribution function takes the Boltzmann form,
\begin{equation}\label{eq:fBoltzmann}
f(E)\propto e^{-E/T_B} \, .
\end{equation}
It is straightforward to check that (within the approximation of neglecting the Pauli blocking factors) this form satisfies the equation
\begin{equation}
    St^{Phon}_{sp}\left\{f\right\} + St^{Phon}_{st}\left\{f,n_T\right\} = 0 \, .
\end{equation}
Then using this equation and Eq.~\eqref{eq:ApproxGamma^Phon_sp}, which is valid irrespective of the exact form of $f$,  we can write
\begin{equation}\label{eq:StPhonst_est}
    St^{Phon}_{st} \simeq St^{Phon}_{sp} \Big|_{E=E_*} \left(\frac{E-\Delta}{E_*-\Delta}\right)^3 e^{-(E-E_*)/T_B}
\end{equation}
Moreover, using Eq.~\eqref{eq:fBoltzmann} the photon collision integral can be written in the form
\begin{equation}
    St^{Phot} = St^{Phot}\Big|_{E=E_*} e^{-(E-E_*)/T_B}
\end{equation}
Since $St^{Phot}|_{E=E_*} = St^{Phon}_{sp}|_{E=E_*}$, comparing the last two equations we see that, in fact, $St^{Phot} < St_{st}^{Phon}$ for $E>E_*$. Note that to get the estimate in Eq.~\eqref{eq:StPhonst_est} we have assumed the validity of Eq.~\eqref{eq:fBoltzmann} at all energies, including $E<E_*$, and this assumption underestimates the contribution to $St_{st}^{Phon}$ from the term in Eq.~\eqref{sec:Rate Equation eq: Stimulated Phonon integral} proportional to $f(E-\omega)$ originating from the interval $E-E_*<\omega<E-\Delta$. However, this contribution is smaller than that coming from the interval $0<\omega<E-E_*$ if $E-E_* \gtrsim E_*-\Delta-\tilde{T}_*(\tilde{T}/T_B)^{2/3}$ [with $E_*$ of Eq.~\eqref{sec: App eq: Approximate E_* High E}], an inequality that identifies the width of the crossover region between the approximate expressions valid below and above $E_*$. These considerations are for the case $E_*>2\Delta$, but they can be extended to the interval $E_* < E < 2\Delta$ in the remaining case.

\section{Nonequilibrium phonon distribution}
\label{appendix:noneqphonons}

Here we study the nonequilibrium form of the phonon distribution function. We start by considering energies below the pair-breaking threshold, $\omega < 2\Delta$. Then writing $n=n_T+n_1$, in the steady state Eq.~\eqref{sec:Rate Equation eq: Phonon Rate Equation} takes the form (neglecting as usual Pauli-blocking factors)
\begin{align}
    & n_1(\omega) = \frac{2\tau_l}{\pi \Delta_0 \tau_0^{PB}}\int\limits_{\Delta}^{\infty}\!dE\, \rho(E) U^-(E,E+\omega) \\ &\big\{ f(E+\omega) 
\left[1+n_T(\omega)+n_1(\omega)\right]  -f(E) \left[n_T(\omega)+n_1(\omega)\right] \! \big\} \nonumber
\end{align}
To solve this equation approximately, we consider two limiting cases. First, we consider energies such that $n_1 \ll n_T$; then we can neglect $n_1$ in the right-hand side, and the resulting expression amounts to the first iterative solution. To this case belongs in particular the low-energy regime $\omega\ll T_B$, in which case we can approximate $n_T(\omega) \simeq T_B/\omega$, $f(E+\omega) \simeq f(E) + \omega f'(E)$, $U^-(E,E+\omega)\simeq U^-(E,E)=1/\rho(E)$ and therefore find $n_1 \simeq (2\tau_l/\pi\tau_0^{PB})(T_B/\Delta_0)b_0$. Then typically $n_1 \ll n_T$ unless $\tau_l$ is several orders of magnitude longer than $\tau_0^{PB}$. 

A second limiting case is when $n_T \ll n_1 \ll f(E+\omega)/f(E)$ (since $f$ is in general monotonically decreasing, the latter inequality also implies $n_1 \ll 1$). In this case $n_1$ is approximately as in Eq.~\eqref{sec:Shape of the quasiparticle distribution eq: Definition n_1} and using Eq.~\eqref{sec: Shape of the quasiparticle distribution eq: Low Energy Solution high energy} we have, for $T_* \lesssim \omega \lesssim \Delta$,
\begin{equation}
    f(E+\omega) \simeq \frac{3b_0}{2^{5/6}\sqrt{\pi}}\left(\frac{E+\omega-\Delta}{T_*}\right)^{-\frac{1}{2}}e^{-[(E+\omega-\Delta)/T_*]^3/3}
\end{equation}
In the argument of the exponential function, we can keep terms linear in $(E-\Delta)/T_*$  while disregarding higher powers, and for $\omega \gtrsim T_*$ we can approximate $U^-(E,E+\omega)\simeq \omega/\sqrt{\omega^2+2\omega\Delta}$. In integrating over energy $E$, we can also approximate $\rho(E)\simeq\sqrt{\Delta/(2(E-\Delta))}$ to arrive at
\begin{equation}\label{eq:n1_explicit}
    n_1(\omega) \simeq \frac{3}{2^{1/3}\pi}b_0 \frac{\tau_l}{\tau_0^{PB}}\left(\frac{T_*}{\Delta}\right)^2 \frac{\Delta}{\omega}\sqrt{\frac{\Delta}{2\Delta+\omega}}
    e^{-(\omega/T_*)^3/3}.
\end{equation}
For $\omega\gtrsim\Delta$ we can proceed as for $\omega\lesssim \Delta$, but using Eq.~\eqref{sec: Shape of the quasiparticle distribution eq: High Energy Solution approximate} instead of Eq.~\eqref{sec: Shape of the quasiparticle distribution eq: Low Energy Solution high energy}; this way we obtain
\begin{align}\label{eq:n1_explicit2}
    n_1(\omega)&\simeq  \frac{2^{1/6}3^{3/2}}{\sqrt{5}\pi}b_0\frac{\tau_l}{\tau_0^{PB}} \left(\frac{T_*}{\Delta}\right)^2 \frac{\Delta}{\omega}\sqrt{\frac{\Delta}{2\Delta+\omega}} \nonumber \\ &\times\exp\left[-\frac{2}{5}\sqrt{\frac{35}{64}}\left(\frac{\omega}{T_*}\right)^{5/2}\sqrt{\frac{\Delta}{T_*}} \right]
\end{align}
Note that the two exponents in Eqs.~\eqref{eq:n1_explicit} and \eqref{eq:n1_explicit2} match at $\omega=63\Delta/80$, which as expected is of order $\Delta$.

To find the crossover frequency $\omega_c$ between thermal distribution at $\omega<\omega_c$ and nonequilibrium distribution at higher energies, assuming $\omega_c \lesssim \Delta$ we equate Eq.~\eqref{eq:n1_explicit} for $n_1$ to $n_T(\omega) \simeq e^{-\omega/T_B}$, which leads to the equation
\begin{align}\label{eq:omegaceq}
    & \frac{1}{3}y^3-\frac{T_*}{T_B}y + \ln y + \frac{1}{2}\ln \left(1+\frac{T_*}{2\Delta}y\right) + \\
    & \ln \left(\frac{2^{5/6}\pi}{3}\frac{\Delta}{T_*}\frac{\tau_0^{PB}}{\tau_l}b_0^{-1}\right) =0 \nonumber
\end{align}
where $y=\omega_c/T_*$. For $y\ll\sqrt{3T_*/T_B}$, we can neglect the first term in this equation, and up to small corrections we arrive at Eq.~\eqref{sec:Shape of the quasiparticle distribution eq: omegac}. More accurate estimates for $\omega_c$ can be found by solving Eq.~\eqref{eq:omegaceq} numerically. For $\omega_c \gtrsim \Delta$ the same approach can be employed, but using Eq.~\eqref{eq:n1_explicit2} for $n_1$.

We now turn to energies above the pair-breaking threshold, $\omega>2\Delta$. Based on the considerations made thus far, so long as $\omega_c \lesssim 2\Delta$ we expect the inequalities $n_T \ll n_1 \ll 1$ to hold, meaning that we could discard $n_T$ in Eq.~\eqref{sec:Rate Equation eq: Phonon Rate Equation}; however, to allow for $\omega_c > 2\Delta$ we keep the $n_T$ term. In both cases, in the second integral in Eq.~\eqref{sec:Rate Equation eq: Phonon Rate Equation} we can neglect $n(\omega)\ll1$ in the first term in curly brackets and, using the approximate energy-independence of $1/\tau_{PB}^{Phon}$ (cf. Appendix~\ref{appendix:Lifetimes}), rewrite the second term as $-n(\omega)/\tau_0^{PB}$. Furthermore, we assume the quasiparticle density to be sufficiently low for $\tau^{Phon}_{abs} \gg \tau^{0}_{PB}$ to hold, so we can neglect the second term in curly brackets in the first integral of Eq.~\eqref{sec:Rate Equation eq: Phonon Rate Equation} (cf. Sec.~\ref{sec:Lifetimes}). Then the approximate solution for the phonon distribution reads
\begin{equation}
\begin{split}\label{eq: Non-Equi-Phonons above Threshold}
   & n(\omega) \simeq \frac{1}{\zeta}\left[n_1(\omega) + n_2(\omega) + n_T(\omega) \right] =\\
   & \frac{\tau_l}{\pi \Delta_0 
    \tau_0^{PB} \zeta}
   \Bigg[2\int\limits_{\Delta}^{\infty}\!dE \, \rho(E) U^-(E,E+\omega) f(E+\omega) \\
    &+\int\limits_{\Delta}^{\omega-\Delta}\!dE \, \rho(E) U^+(E,\omega-E) f(\omega-E)f(E) \Bigg] \\
    & + \frac{ 1}{\zeta} n_T(\omega)
\end{split}
\end{equation}
where
\begin{equation}
    \zeta = 1+\tau_l/\tau_0^{PB}
\end{equation}
is the phonon trapping factor [cf. Eq.~\eqref{eq:tau0bar}]. The first term in the square brackets
coincides with $n_1$ of Eq.~\eqref{sec:Shape of the quasiparticle distribution eq: Definition n_1}, but here the prefactor $1/\zeta$ 
is smaller than unity, since in addition to scattering a new relaxation channel for phonons (that is, pair breaking) is now available. The second term in square brackets, $n_2(\omega)$, takes into account phonon generation by quasiparticle recombination. Assuming the (envelope of the) quasiparticle distribution to be monotonically decreasing, we can bound this term by $b_0^2 \Delta S_+(\omega/\Delta)$ [with $S_+$ of Eq.~\eqref{eq: Spectral Density S+}], which near $\omega=2\Delta$ is approximately $b_0^2\Delta\pi$; then, using Eq.~\eqref{eq:n1_explicit2} (with the appropriate prefactor) we find that the 
first term always dominates over the second one at the threshold $\omega=2\Delta$ if 
\begin{equation}\label{eq:b_0n2condition}
    b_0 < (3\times 2^{1/6}/\sqrt{5}\pi) \zeta (T_*/2\Delta)^2 e^{-\sqrt{35/32}(2\Delta/T_*)^3/5}
\end{equation}
Even if $n_2$ dominates, as discussed in Sec.~\ref{sec:Shape of the quasiparticle distribution Phonon trapping}, these nonequilibrium phonons due to quasiparticle recombination could affect the shape of the quasiparticle distribution only at energies $E>3\Delta$ by contributing to the collision integral $St_{st}^{Phon}$, Eq.~\eqref{sec:Rate Equation eq: Stimulated Phonon integral}.
For $\omega-2\Delta\lesssim T_*$, one can apply the low energy approximation to the product of density of states times coherence factor and use 
Eq.~\eqref{sec: Shape of the quasiparticle distribution eq: Low Energy Solution low energy} to evaluate approximately the second term in the square brackets of Eq.~\eqref{eq: Non-Equi-Phonons above Threshold} to find
\begin{equation}
    n_2(\omega)\approx \frac{b_0^2\tau_l}{\pi (\tau_0^{PB}+\tau_l)}\left[\pi -1.20 (\omega-2\Delta)^{5/2} \right]
\end{equation}
We do not consider here the behavior of $n_2$ for frequencies $\omega>2\Delta+T_*$.

\section{Normalization equation}
\label{appendix: Neglected terms in the Normalization}

We derive here explicit expression for the coefficients $I_i$ ($i=0,\,1,\,2$) entering Eq.~\eqref{eq:Normalization eq} for the normalization constant $b_0$. To find those coefficient, we substitute Eq.~\eqref{eq:nnoneq2D} into Eq.~\eqref{eq:normalization} and divide the result by $(2\Delta)^3/(1+\tau_l/\tau_0^{PB})$. From the term containing $n_T$ we find
\begin{equation}\label{eq: I_0}
    I_0 = \frac{1}{(2\Delta)^3}\int_{2\Delta} d\omega \, \omega^2 n_T(\omega,T_B) \simeq \frac{T_B}{2\Delta} e^{-2\Delta/T_B}
\end{equation}

To find $I_2$, we collect together the terms quadratic in $f$, switch the integration order between $\omega$ and $E$, and change variable from $\omega$ to $E'= \omega-E$; we obtain
\begin{equation}\label{eq: I_2}
\begin{split}
    I_2 = \frac{1}{b_0^2 \pi \Delta (2\Delta)^3} \int_\Delta dE \int_\Delta dE' \,
    (E+E')^2
    \\ \rho(E)U^+(E,E')f(E)f(E')
\end{split}
\end{equation}
The presence of the two distribution functions implies that the main contributions to the integrals come from regions close to $\Delta$, and employing similar approximations as those used in Eq.~\eqref{eq:Nqp} we arrive at
\begin{equation}
    I_2 \simeq \frac{(2.1)^2}{\pi}\frac{T_*}{2\Delta}
\end{equation}

Finally, considering the $n_1$ term we get
\begin{equation}\label{eq: I_1}
    I_1 = \frac{1}{b_0\pi \Delta (2\Delta)^3}\! \int\limits_\Delta\! dE\! \int\limits_{2\Delta} \!d\omega\,\omega^2 \rho(E) U^-(E,E+\omega) f(E+\omega)
\end{equation}
Here we have to consider separately various regimes. For high phonon temperature/low photon number, $T_B^* \lesssim T_B \ll T_*$, the distribution function takes the Boltzmann form $f(E) = b_T e^{-E/T_B}$ above the energy $E_* \lesssim 2\Delta < 3\Delta$, see Sec.~\ref{sec:finteTphonons}, and we have
\begin{equation}\label{eq:I1Boltzmann}
    I_1 \simeq \frac{1}{\sqrt{2\pi}}\frac{b_T}{b_0}\left(\frac{T_B}{2\Delta}\right)^{3/2}e^{-3\Delta/T_B}
\end{equation}
where
\begin{equation}
    \frac{b_T}{b_0} \approx \frac{3}{2^{5/6}\sqrt{\pi}}\left(\frac{T_B}{T_*}\right)^{1/4} e^{\Delta/T_B}e^{(2/3)(T_*/T_B)^{3/2}}
\end{equation}
With these expression, using that $(T_*/\Delta)^3 \lesssim T_B/\Delta$ the condition for the normalization constant $b_0$ to give the thermal equilibrium density can be written as $0.16(\tau_l/\tau_0^{PB})^2 (T_B/\Delta)^2 e^{-2\Delta/3T_B} < 1$; even at the relatively high temperature $T_B = 0.3 T_c$, this condition becomes $10^{-4}(\tau_l/\tau_0^{PB})^2 <1$, showing that in this regime deviations from the thermal equilibrium density are possible only if $\tau_l$ exceeds $\tau_0^{PB}$ by at least a few orders of magnitude. We do not pursue the analysis of this long-$\tau_l$ limit here, as we focus on the experimentally relevant case of aluminum films in which $\tau_l$ is comparable to $\tau_0^{PB}$. However, the long-$\tau_l$ regime could be relevant for other materials, such as niobium.

For low phonon temperature/high photon number, $T_B \ll T_B^*$, we have $E_* \gtrsim 2\Delta$, so we need to distinguish between $E_* < 3\Delta$ and $E_* > 3\Delta$. In the former case, Eq.~\eqref{eq:I1Boltzmann} still holds, but now
\begin{equation}\label{eq:btb0_lowphonon}
    \frac{b_T}{b_0} \approx \frac{3^{3/2}}{2^{1/3}\sqrt{5\pi}}\sqrt{\frac{35}{64}} \left(\frac{T_B}{\Delta}\right)^{1/2}\left(\frac{T_*}{\Delta}\right)^{-1}e^{\frac{\Delta}{T_B}}e^{ \frac35\left(\frac{\tilde{T}_*}{T_B}\right)^{5/3}}
\end{equation}
Using this expression, one can check that the condition for $b_0$ to give the thermal equilibrium density is easily violated, due to the last exponential factor in Eq.~\eqref{eq:btb0_lowphonon} being dominant and large in the regime $T_B/\Delta \ll (T_*/\Delta)^3$ (the violation takes place unless $\tau_l/\tau_0$ is small compared to the inverse of that exponential factor). Similarly, in the case $E_*>3\Delta$ we find again that the density generically deviates from the thermal equilibrium one. In this case we have
\begin{equation}\label{eq:I1highphoton}
    I_1=\frac{3^{3/2}}{2^{7/3}5\pi\sqrt{7}}\left(\frac{T_*}{\Delta}\right)^5\exp\left[-\sqrt{\frac{14}{5}}\left(\frac{\Delta}{T_*}\right)^3\right]
\end{equation}
and in the ratio $I_1^2/(I_0 I_2)$ the exponential factor $e^{2\Delta/T_B}$ from $I_0$ dominates over the exponential in $I_1$. Note that despite the similarity, the coefficient $I_1$ and hence the value of $b_0$ in the two cases are different. Also, although we have included the case $E_*<3\Delta$ for completeness, it has limited relevance, since the inequality holds if $(T_*/\Delta)^3 < 2^{3/2}\sqrt{35/64}\left(T_B/\Delta\right)$, and at the same time we require $T_B/\Delta \ll (T_*/\Delta)^3$.

To summarize, we have found that for high phonon temperature/low photon number the quasiparticle density is the same as in thermal equilibrium and the term linear in $b_0$ in Eq.~\eqref{eq:Normalization eq} can be neglected. Conversely, at low phonon temperature/high photon number the constant term can be neglected. The relationship to the generalized Rothwarf-Taylor equation can be found by restoring all prefactors; in practice, this amounts to expressing $b_0$ in terms of $N_\mathrm{qp}$ using Eq.~\eqref{eq:Nqp}, multiplying the left-hand side of Eq.~\eqref{eq:Normalization eq} by $4\rho_F(2\Delta/T_c)^3 \pi \Delta/\bar{\tau}_0$, and equating the result to $dN_\mathrm{qp}/dt$. In this procedure, we can use for $I_1$ the formula in Eq.~\eqref{eq:I1highphoton}; in this way, the crossover between the two regimes is correctly identified up to numerical factors of order unity.

\section{Photon-number dependent corrections in the generalized RT model}
\label{appendix: Higher Order Corrections to the Quasiparticle density}

In the discussion of the generalized RT model in Sec.~\ref{sec:genRT} we limited our considerations to the leading order in the parameter $T_*/\Delta \lesssim 1$.  However, corrections in powers of $\varepsilon \equiv T_*/\Delta$ can be taken into account, as we now show. We begin by evaluating corrections to the expression for the quasiparticle density, Eq.~\eqref{eq:Nqp}. In the integral there, we make the change of variables $E=\Delta(1+\varepsilon x)$ and express the density of states as $\rho(E) \simeq \sqrt{\Delta/2T_* x}\left(1+3\varepsilon x/4 - 5(\varepsilon x)^2/32\right)$. Defining $a_\beta=\int_0 dx \, x^\beta f(x)/b_0$, we find 
\begin{equation}\label{eq:Nqp_corr}
    N_\mathrm{qp} \simeq 2\rho_F \sqrt{2 T_* \Delta}\,b_0 \left(a_{-1/2} + \frac{3}{4}a_{1/2} \varepsilon - \frac{5}{32} a_{3/2} \varepsilon^2 \right)
\end{equation}
The $a_\beta$ factors can be estimated numerically; they have the approximate values $a_{-1/2}\simeq 2.1$, $a_{1/2} \simeq 0.88$, and $a_{3/2}\simeq 0.77$.

\begin{figure}
    \centering
    \includegraphics[scale=.5]{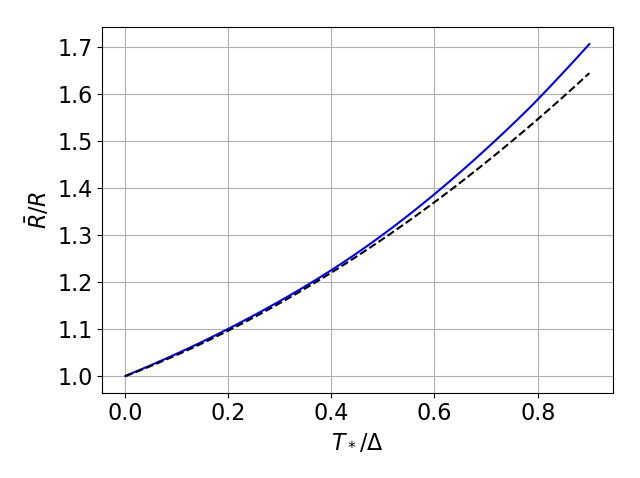}
    \caption{Blue (solid): numerically calculated recombination coefficient for the distributions derived in Sec.~\ref{sec: Shape of the quasiparticle distribution} (for $T_B=0$), in units of the (zeroth-order) recombination coefficient $R$ defined in Eq.~\eqref{eq:Rdef}. Black (dashed): analytical approximation  Eq.~\eqref{eq:barR}.}
    \label{fig:R_bar}
\end{figure}

The same procedure can be applied to the evaluation of $I_2$ of Eq.~\eqref{eq: I_2} by expanding in the integral the energy-dependent factors multiplying the two distribution functions. By comparing the result to the square of Eq.~\eqref{eq:Nqp_corr}, we find that the recombination term in Eq.~\eqref{eq:genRT} can be written as $\bar{R} N_\mathrm{qp}^2$ with
\begin{equation}\label{eq:barR}
    \frac{\bar{R}}{R} =  1+\frac{a_{1/2}}{a_{-1/2}}\varepsilon + \left[\frac{5}{4}\frac{a_{3/2}}{a_{-1/2}} - \frac{3}{4}\left(\frac{a_{1/2}}{a_{-1/2}}\right)^2\right]\varepsilon^2
\end{equation}
and $R$ defined in Eq.~\eqref{eq:Rdef}. In Fig.~\ref{fig:R_bar} we show that this second-order-in-$\varepsilon$ expression gives a reasonable approximation for the recombination coefficient even for $T_*$ close to $\Delta$.

Additional corrections originate from taking into account the dependence on $\omega$ of the phonon pair-breaking lifetime $\tau_{PB}^{Phon}(\omega)$, Eq.~\eqref{eq:tPBPhonS}, which we have previously neglected. In the expression for the phonon distribution function Eq.~\eqref{eq: Non-Equi-Phonons above Threshold}, this amounts to the substitution $\zeta \to \zeta(\omega)$, where the energy-dependent phonon trapping factor is
\begin{equation}
    \zeta(\omega) = 1+\tau_l/\tau_{PB}^{Phon}(\omega) \, .
\end{equation}
We also need to modify Eq.~\eqref{eq:normalization} by multiplying the distribution function $n(\omega)$ by $\tau_0^{PB}/\tau_{PB}^{Phon}(\omega)$. Inserting the thus corrected Eq.~\eqref{eq: Non-Equi-Phonons above Threshold} into the modified Eq.~\eqref{eq:normalization}, we find that the two integrals $I_0$ and $I_1$ for the generation terms are modified by inserting in the integrands of Eqs.~\eqref{eq: I_0} and \eqref{eq: I_1} a factor $\tau_0^{PB} \zeta/\tau_{PB}^{Phon}(\omega)\zeta(\omega)$, while in the integrand for the generation term $I_2$, Eq.~\eqref{eq: I_2}, we must include the factor $\zeta/\zeta(E+E')$. Since in the relevant regime ($T_B \ll T_B^*$) $I_1$ has stronger than exponential dependence on $T_*/\Delta$, see Eq.~\eqref{eq:I1highphoton}, we do not pursue the calculation of weak corrections proportional to $\varepsilon^{9/5}$ [in fact, the main correction to the function $G(x)$ originates from the linear term in Eq.~\eqref{eq:Nqp_corr} for $N_\mathrm{qp}$, so that the right-hand side of Eq.~\eqref{eq:Gdef} should be multiplied by $1-(3a_{1/2}/4a_{-1/2})\varepsilon$]. Similarly, we neglect the small corrections in $T_B/\Delta$ that would be introduced to the leftmost expression in Eq.~\eqref{eq: I_0}. For $I_2$, we limit ourselves to terms linear in $\varepsilon$; at this order we find that the coefficient of the linear term in Eq.~\eqref{eq:barR} should be multiplied by $(1+\tau_l/2\tau_0^{PB})/(1+\tau_l/\tau_0^{PB})$. As function of $\tau_l$, this factor varies between 1 and 1/2, so the finite thermalization time can weaken the dependence of the recombination coefficient $\bar{R}$ on $T_*/\Delta$, but it does not change its increase with this parameter. Therefore, in the high phonon temperature regime $T_B \gtrsim T_B^*$ in which the $I_1$ term can be neglected, this increase implies a decrease of quasiparticle number with increasing photon number.

\section{Parameters for comparison to experiment}
\label{appendix: Simulation Parameter}

We discuss here our estimates for the parameters used in the comparison to the experiment of Ref.~\cite{Visser.2014}, see Table~\ref{tab: Param Ex} in Sec.~\ref{sec:expcomp}. The kinetic inductance fraction $\alpha$ and the zero-temperature gap $\Delta_0$ are obtained by fitting the measured internal quality factor for temperatures $T>0.25\,$K and the lowest readout power, $P_\mathrm{read}=-100\,$dBm, to the thermal equilibrium expression~\cite{Gao.2008}
\begin{equation}
    Q_i=\frac{\pi  }{4 \alpha\sinh(x)K_0(x)}\exp \left(\frac{\Delta_T}{T}\right)
\end{equation}
where $x= \omega_0/(2T)$ and $K_0$ is the 0-th order modified Bessel function. A least square fit gives $\alpha\simeq 0.13$ and $\Delta_0\simeq 189\,\mu$eV. The assumption of the thermal equilibrium for the quasiparticles is justified, since the estimated $T_*$ is comparable to $\omega_0$, see Table~\ref{tab:Limiting Quality factors}.

The kinetic inductance fraction can be estimated based on the geometry of the resonator using Eqs.~(8) and (46) in Ref.~\cite{Clem.2013} (in the latter equation, the penetration depth can be estimated using the measured value of resisitivity); this yields $\alpha = 0.07$, which is of the same order as the value obtained from the fit. In Ref.~\cite{Visser.2014} a smaller value for $\Delta_0$ is given, estimated by measuring the critical temperature and using the BCS relation for the ratio $\Delta_0/T_c$; our estimate agrees with previous findings of a higher ratio in thin aluminum films~\cite{Chubov1969}.
To estimate $c^{QP}_{Phot}$ using Eq.~\eqref{Resonator_Q}, we follow Ref.~\cite{Visser.2014} and assume the volume occupied by the quasiparticles to be twice the central strip volume, $V=2\cdot 1770\,\mu$m$^3$, to account for quasiparticles in the ground plane, and take $\rho_F=1.74\times 10^4/\mu$eV$\mu\mathrm{m}^3$. Then we use that for thin-film resonators we have $Q'=\pi \Delta_0/\alpha \omega_0$, as can be seen by replacing $\sigma_1 \rightarrow \sigma_N$ in the central expression in Eq.~\eqref{sec: Quality factors eq: Q_i 1}.

\bibliography{Sources}

\end{document}